\documentclass[preprint,aps,prd,groupedaddress,showpacs,nofootinbib]{revtex4-1}%
\usepackage{graphicx}
\usepackage{amsmath}
\usepackage{amssymb}
\usepackage{bm}
\usepackage{epsfig}
\usepackage{dcolumn}
\usepackage{slashed}
\usepackage{empheq}

\def\taun{{\cal T}_N}
\newcommand{\dotp}[2]{#1 \cdot #2}

\newcommand{\bea}{\begin{eqnarray}}
\newcommand{\eea}{\end{eqnarray}}
\newcommand{\nn}{\nonumber}

\begin{document}

\setlength\baselineskip{17pt}

\title{The $N$-jettiness soft function at next-to-next-to-leading order}

\author{Radja Boughezal}
\email{rboughezal@anl.gov}
\affiliation{High Energy Physics Division, Argonne National Laboratory, Argonne, IL 60439,USA}

\author{Xiaohui Liu}
\email{xhliu@umd.edu}
\affiliation{Maryland Center for Fundamental Physics, University of Maryland, College Park, Maryland 20742, USA} 
\affiliation{Center for High-Energy Physics, Peking University, Beijing, 100871, China}

\author{Frank Petriello}
\email{f-petriello@northwestern.edu}
\affiliation{Department of Physics \& Astronomy,Northwestern University, Evanston, IL 60201,USA}
\affiliation{High Energy Physics Division, Argonne National Laboratory, Argonne, IL 60439,USA}

\date{\today}

\begin{abstract}
We present a general framework for the calculation of soft functions for SCET$_{\text{I}}$ observables through next-to-next-to-leading order (NNLO) in the strong coupling constant.  As an example of our formalism we show how it can be used to obtain the complete NNLO soft function for the $N$-jettiness event shape variable.  We present numerical results for two examples with phenomenological impact: the one-jettiness soft function for both electron-proton and proton-proton collisions.

\end{abstract}

\maketitle

\section{Introduction}

The physics program at the Large Hadron Collider (LHC) and at other experiments increasingly relies upon precision calculations within the Standard Model (SM) in order to search for small deviations indicative of new physics.  At the LHC, Run I was marked by the discovery and initial characterization of the Higgs boson.  Run II will focus on the detailed investigation of this new state, in which an evermore precise characterization of the SM benchmark will be critical.  The small errors for most experimental measurements make higher-order QCD calculations mandatory in interpreting the data.  Such computations may be performed at either fixed-order in QCD perturbation theory, or may additionally include the resummation of large logarithms in certain regions of phase space, either through analytic resummation or via the use of parton-shower simulations.

One feature of recent progress in precision calculations is the impact of analytic resummation techniques on our understanding of jet properties.  They led to the invention of new variables such as $N$-jettiness~\cite{Stewart:2010tn} and $N$-subjettiness~\cite{Thaler:2010tr} that describe jet substructure.  Additionally, resummation of large logarithmic corrections improved our description of the theoretical treatment of Higgs production in exclusive jet bins~\cite{Stewart:2013faa,Banfi:2012yh, Banfi:2012jm,Becher:2012qa,Banfi:2013eda,Becher:2013xia,Tackmann:2012bt,Gangal:2014qda,Liu:2012sz,Liu:2013hba,Boughezal:2013oha,Boughezal:2014qsa,Boughezal:2015oga}.  The starting point for analytic resummation is a factorization theorem describing the observable under consideration, usually in a region of phase space where an expansion of the full QCD result is possible.  A typical factorized cross section takes on the schematic form
\begin{equation}
 \sigma \sim  \int H \otimes B \otimes B \otimes S \otimes   \left[ \prod_n J_n \right] .
\end{equation}
Here, $H$ describes the effect of hard radiation, $B$ encodes the effect of radiation collinear to one of the two initial beam directions, $S$ describes the soft radiation, and $J_n$ contains the radiation collinear to a final-state jet.  Depending on the observable and process under consideration, only a subset of these terms may be present.  Perturbative corrections to each of these functions are minimized by the appropriate renormalization scale choice.  Renormalization group equations for each separate function allow these scales to be evolved to a common one, in the process resumming large logarithms.  We have used the language of soft-collinear effective theory (SCET)~\cite{Bauer:2000ew,Bauer:2000yr,Bauer:2001ct,Bauer:2001yt,Bauer:2002nz} in our description, although similar quantities appear in other approaches to resummation.

Improving the accuracy of resummation requires both knowledge of the anomalous dimensions controlling the evolution of the various functions in the factorization theorem, and the perturbative expansion of these quantities to higher orders in the strong coupling constant $\alpha_s$.  Knowledge of the singular structure of QCD gained with resummation formulae has also improved our ability to calculate fixed-order QCD quantities to higher precision.  A well-known example is the use of $q_T$-subtraction to calculate cross sections through next-to-next-to-leading order (NNLO)~\cite{Catani:2007vq}.  Currently, the hard function $H$ is known to NNLO for numerous phenomenologically interesting processes containing final-state jets~\cite{Bern:2000dn,Anastasiou:2000kg,Anastasiou:2000ue,Anastasiou:2001sv,Gehrmann:2011aa,Gehrmann:2011ab}.  The NNLO beam functions for observables such as jettiness and beam thrust are available~\cite{Gaunt:2014xga,Gaunt:2014cfa}, as are the NNLO final-state jet functions~\cite{Becher:2006qw,Becher:2010pd}.  The only missing ingredient of the factorization formula at NNLO for observables such as jettiness and beam thrust is the  soft function.  The knowledge of these soft functions do not only enable the resummation accuracy of the corresponding observables to be improved; they are also important components of the recently proposed jettiness-subtraction scheme for the calculation of jet cross sections through NNLO in QCD~\cite{wjet}.

We present in this manuscript a general method of computing soft functions through NNLO for SCET$_{\text I}$ observables.  It uses sector decomposition~\cite{Hepp:1966eg,Binoth:2000ps,Anastasiou:2003gr} to extract singularities from the integrals which occur in the calculation and reduce them to a form amenable to numerical integration.  We illustrate our techniques using the $N$-jettiness event shape variable ${\cal T}_N$ in as an example.  We validate our approach using against known results in the literature.  The $N$-jettiness soft function contains the logarithms $\text{ln} ({\cal T}_N)$, and a contribution of the form $\delta ({\cal T}_N )$.  The logarithmic corrections at NNLO can be obtained by expanding the resummed expression for the soft function to ${\cal O}(\alpha_s^2)$.  We demonstrate that we reproduce these known results with our technique.  Our computation of the $\delta ({\cal T}_N )$ correction is new.  We present numerical results for two selected examples of recent phenomenological relevant: the one-jettiness soft function in electron-proton collisions, and the one-jettiness soft function in proton-proton collisions.

Our manuscript is organized as follows.  We review the definition of the $N$-jettiness event shape variable in Section~\ref{sec:jettiness}.  Our calculational framework is presented in Section~\ref{sec:framework}, where we show how to reduce the NNLO soft function to a form suitable for numerical evaluation.  In Section~\ref{sec:numerics} we present numerical results for two examples: one-jettiness in electron-proton collisions, and one-jettiness in proton-proton collisions.  We conclude in Section~\ref{sec:conc}.

\section{Description of jettiness}\label{sec:jettiness}

We begin with a brief review of the $N$-jettiness event-shape variable ${\cal T}_N$ of Ref.~\cite{Stewart:2010tn}.  ${\cal T}_N$ is defined by
\begin{equation}
{\cal T}_N = \sum_k \text{min}_i \left\{ \frac{2 p_i \cdot q_k}{Q_i}\right\}.
\end{equation}
Here, the $p_i$ are light-like reference vectors for each of the initial beams and final-state jets in the problem, while the $q_k$ denotes the four-momentum of final-state radiation radiation. The $Q_i$ are dimensionful variables that characterize the hardness of the beam-jets and final-state jets.  For simplicity, we will set $Q_i = 2 E_i$, twice the energy of each jet.  Writing the jet momenta as $p_i = E_i n_i$, we have 
\begin{equation}
{\cal T}_N = \sum_{k} \text{min}_{i} \left\{ n_{i} \cdot q_{k}\right\}.
\end{equation}
We will consider the calculation of $\taun$ through NNLO in QCD, which will receive contributions from single emission and double emission processes.  The contributions of single and double-real emission processes to $N$-jettiness are given explicitly by the following expressions:
\begin{itemize}

\item Single-real emission: $\taun = \text{min}_i \left\{ n_i \cdot q_1 \right\}$;

\item Double-real emission: $\taun = \text{min}_{i} \left\{ n_i \cdot q_1 \right\} + 
	\text{min}_j \left\{ n_j \cdot q_2\right\}$.

\end{itemize}

\begin{figure}[!h]
\begin{center}
\includegraphics[width=0.40\textwidth,angle=0]{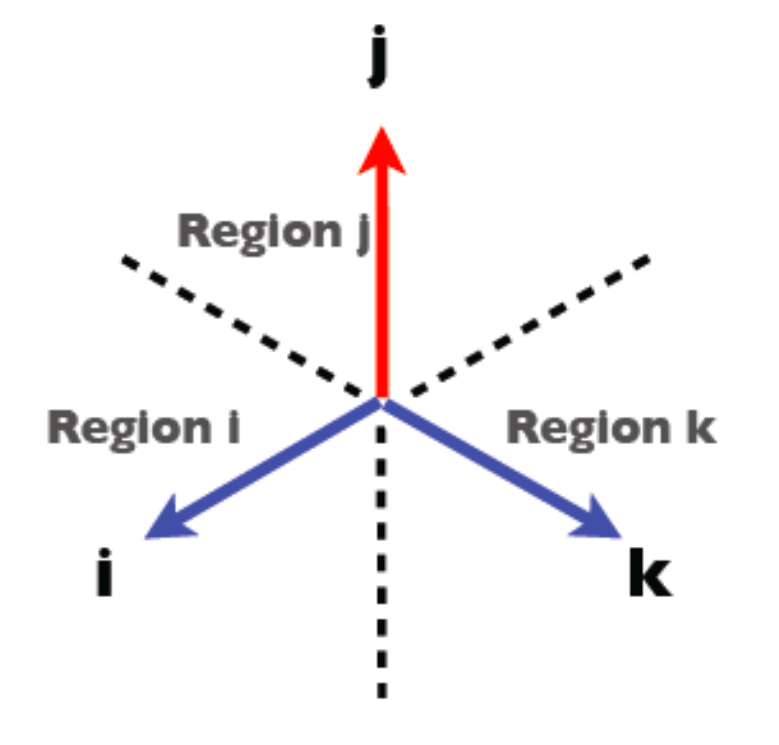}
\end{center}
\vspace{-0.5cm}
\caption{Sketch of the division of phase space into regions for the one-jettiness variable.  $i$, $j$  and $k$ denote representative hard directions.  A two-dimensional projection of the full space has been performed for simplicity of presentation.} \label{fig:regions}
\end{figure}

The physical content of the above expression is that jettiness partitions the phase space of each emission according to which external direction it is nearest.  A pictorial representation of this is given in Fig.~\ref{fig:regions}.  In each of the regions, $\taun$ is defined differently in terms of the radiation four-momentum.  This leads to the insertion of the following measurement functions into the phase space for single and double-real emission processes: 
\begin{equation}
{\cal M}(q_1) = \sum_{i=1}^{N} \Theta_1^i, \;\;\; {\cal M}(q_1,q_2) = \sum_{i,j=1}^{N} \Theta_{12}^{ij},
\label{eq:measurementfunc}	
\end{equation}
where we have abbreviated
\begin{eqnarray}
\Theta_i^r &=& \delta(\taun-n_r \cdot q_i) \prod_{k \neq r} \theta(n_k \cdot q_i-n_r \cdot q_i), \nonumber \\
\Theta_{ij}^{rs} &=& \delta(\taun-n_r \cdot q_i-n_s \cdot q_j) \prod_{k \neq r} \theta(n_k \cdot q_i-n_r \cdot q_i) \prod_{l \neq s} \theta(n_l \cdot q_j-n_s \cdot q_j) .
\end{eqnarray}
For the QCD processes considered here, $\Theta_{ij}^{rs}$ is symmetric under interchange of either its upper or lower indices.  This allows us to reduce the number of phase-space regions relevant for the calculation of double-real emission processes from nine to six.

\section{Calculational Framework} \label{sec:framework}

With the variable $\taun$ and the structure of the measurement function discussed, we are now ready to discuss the calculation of the soft function.  The soft function can be expanded as a perturbation series in the strong coupling constant,
\begin{equation}
S(\taun) = S^{(0)}(\taun) + \frac{\alpha_s}{2\pi}  S^{(1)}(\taun) + \left(\frac{\alpha_s}{2\pi}\right)^2  S^{(2)}(\taun).
\end{equation}
We have suppressed the dependence on the renormalization scale $\mu$.  The leading-order result $S^{(0)}(\tau)$ is just $\delta(\taun)$, while the calculation of the NLO contribution $S^{(1)}(\taun)$ has been discussed extensively in Ref.~\cite{Jouttenus:2011wh}.  We focus our attention on the computation of $S^{(2)}(\taun)$.  The diagrammatic contributions to the integrand involve emission of gluons from eikonal lines, but are most easily obtained from known results for the NNLO soft limits of QCD amplitudes.  As with all NNLO calculations there are contributions from two-loop virtual corrections, one-loop virtual corrections to single-real emission processes (real-virtual), and double-real emission diagrams.  The two-loop virtual corrections are scaleless in dimensional regularization, leaving only the real-virtual and double-real corrections.  We are left with the following pieces to calculate:
\begin{itemize}

\item the real-virtual corrections to the single-gluon emission process;

\item the $q\bar{q}$ double-real emission correction;

\item the double-real gluon emission contribution.

\end{itemize}

\subsection{The real-virtual correction}

\begin{figure}[!h]
\begin{center}
\includegraphics[width=0.55\textwidth,angle=0]{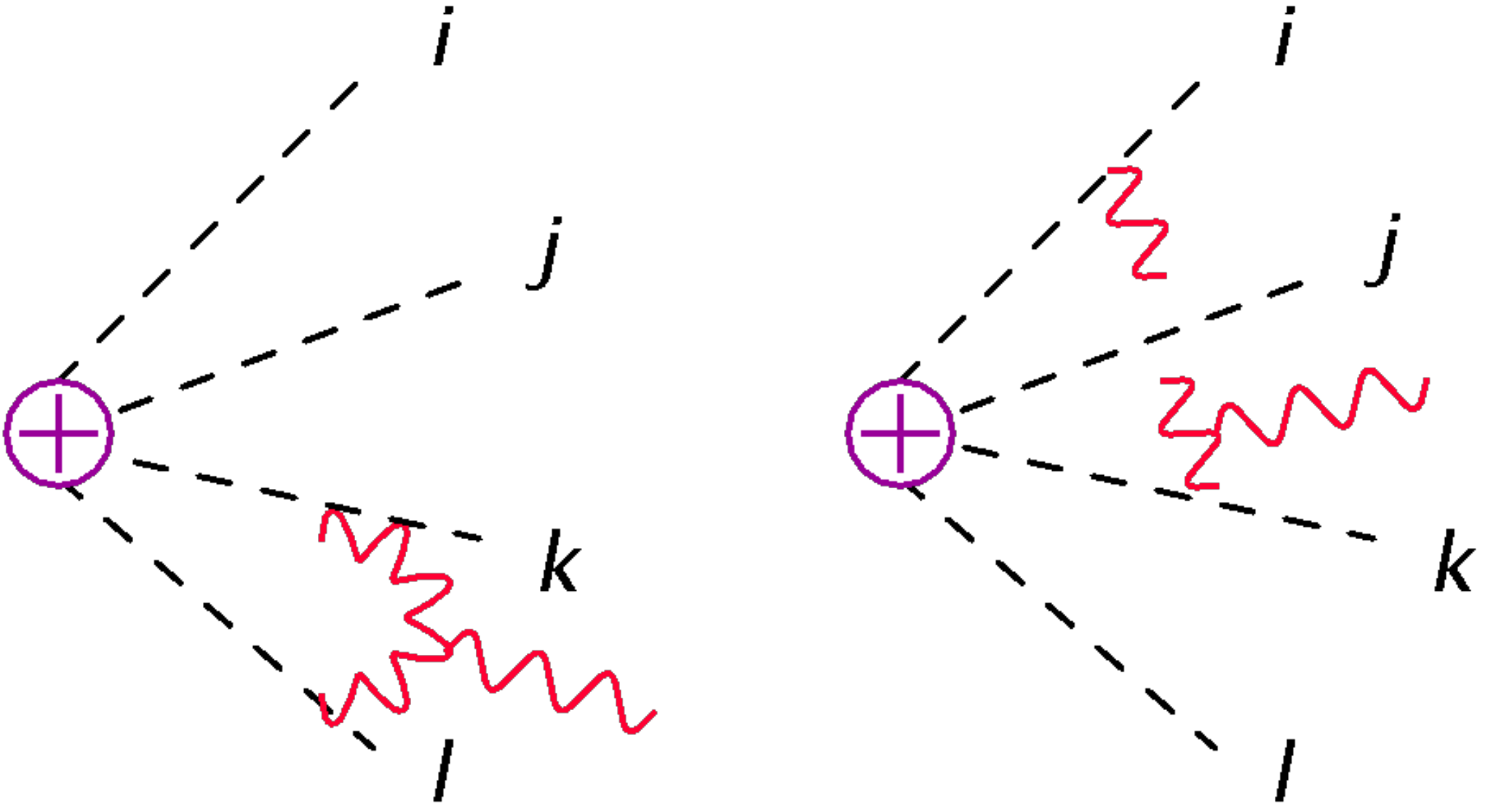}
\end{center}
\vspace{-0.5cm}
\caption{Representative diagrams contributing to the real-virtual piece of the soft function.  The dashed lines represent the eikonal directions.} \label{fig:RV}
\end{figure}

We begin by discussing the real-virtual contribution to the soft function.  It receives contributions from diagrams of the form shown in Fig.~\ref{fig:RV}.  The integrand resulting from these diagrams can be obtained from the one-loop soft-gluon current in QCD~\cite{Catani:2000pi}. We write the real-virtual part of the soft function as
\begin{equation}
S^{(2)}_{RV}(\taun) = \int \frac{d^{d-1}q_1}{(2\pi)^{d-1}} \, E^{(2)}_{RV}(q_1) \, {\cal M}(q_1),
\end{equation}
where the explicit form of the integrand $E^{(2)}_{RV}(q_1)$ can be obtained from Eq.~(26) of Ref.~\cite{Catani:2000pi}.  Although the general structure of this expression is complex, the structure of QCD at NNLO guarantees that the result takes the form of a sum of emissions of $q_1$ from a dipole pair $(i,j)$, where $i,j$ denote two hard directions in the problem, together with appropriate color correlations.  We are therefore led to consider the integration of the following building blocks from which the real-virtual corrections for $\taun$ can be constructed:
\begin{eqnarray}
 {\cal I}_{ij}(q_1) &=& -\frac{8\pi^2 \,C_A}{\epsilon^2} e^{2\epsilon \gamma_E}(4\pi)^{-\epsilon} \frac{\Gamma^4(1-\epsilon)\Gamma^3(1+\epsilon)}{\Gamma^2(1-2\epsilon)
 	\Gamma(1+2\epsilon)} \, [S_{ij}(q_1)]^{1+\epsilon}, \nonumber \\
S_{ij}(q_1) &=& \frac{n_i \cdot n_j}{ 2 \, n_i \cdot q_1 \, n_j \cdot q_1}	.
\end{eqnarray}
$C_A=3$ is the usual QCD color constant.  $\gamma_E$ is the Euler constant, which arises from rewriting the bare coupling constant in terms of the renormalized one.  Since the building block for the soft function is the integral of the ${\cal I}_{ij}(q_1)$ over the real-emission phase space, we will consider the auxiliary quantity
\begin{equation}
I^{(2),ij}_{RV}(\taun) = \int \frac{d^{d-1}q_1}{(2\pi)^{d-1}} \, {\cal I}_{ij}(q_1)\, {\cal M}(q_1),
\end{equation}
from which we can form the entire integrated real-virtual correction.

We now discuss the appropriate representation of phase space for the calculation of this integral.  We introduce a Sudakov decomposition of the radiation momentum $q_1$ in terms of two light-like directions $n_m$ and $n_n$:
\begin{equation} \label{eq:Suddecomp}
q_1^{\mu} = q_1^{+} \frac{n_m^{\mu}}{n_m \cdot n_n } + q_1^{-} \frac{n_n^{\mu}}{n_m \cdot n_n } +q_{1\perp}^{\mu}
\end{equation}
with $q_1^{+} = n_n \cdot q_1$, $q_1^{-} = n_m \cdot q_1$, and $n_m \cdot q_{1\perp} = n_n \cdot q_{1\perp} = 0$.  With this decomposition it is straightforward to write the following expression for the phase space:
\begin{equation}
\int \frac{d^{d-1}q_1}{(2\pi)^{d-1}} = \frac{\Omega_{d-3}}{4 (2\pi)^{d-1}} \frac{1}{n_m \cdot n_n} \left( \frac{n_m \cdot n_n}{2}\right)^{\epsilon} 
	\taun^{2-2\epsilon} \int d\xi \, ds \, d\phi_1 \, \xi^{1-2\epsilon} s^{-2+\epsilon} \text{sin}^{-2\epsilon}(\phi_1),
\end{equation}
where we have set $q^+ = \taun \xi$, $q^- = \taun \xi/s$, and have used the angle $\phi_1$ to parameterize the orientation of $q_{1\perp}$ in the azimuthal plane.  

Upon plugging in the measurement function of Eq.~(\ref{eq:measurementfunc}) and the integrand ${\cal I}_{ij}(q_1)$ into the phase space, we arrive at three terms, depending on which $\Theta_1^{r}$ occurs.  These form two distinct sets: two integrals in which $r$ is one of the two directions $i,j$ appearing in the integrand ${\cal I}_{ij}$, and one in which it does not.  We consider the two integrals corresponding to $r=i$ and $r=k$ with $k$ distinct from $i,j$.  The $r=i$ case can be written as
\begin{equation}
I^{(2),ij}_{RV,i}(\taun) = \int \frac{d^{d-1}q_1}{(2\pi)^{d-1}} \, {\cal I}_{ij}(q_1)\, \Theta_1^{i},
\end{equation}
where we have introduced the subscript $i$ to denote this contribution.  The case $r=j$ can be obtained by simply permuting the indices $i$ and $j$ in this result.  It is convenient to choose the light-cone directions $n_m = n_i$, $n_n = n_j$ for this integral.  Doing so, it is straightforward to derive the following final expression for the integral:
\begin{equation}
\begin{split}
I^{(2),ij}_{RV,i}(\taun) &= -\frac{C_A}{\epsilon^2} B_{RV} \, \taun^{-1-4\epsilon} ( n_i \cdot n_j)^{2\epsilon} \, 2^{-1-4\epsilon} \int_0^1 ds \, dx_2 \, 
	s^{-1+2\epsilon} \,  \text{sin}^{-2\epsilon}(\phi_1)  \\ & \times \prod_{k \neq i,j} \theta\left[ A_{ij,k}(s,\phi_{1k}) -s \right].
\end{split}
\end{equation}
We have set $\phi_1 = 2\pi x_2$, and have introduced the angles $\phi_{1k}$ that denote the separation between $q_1$ and the hard directions $k$ in the transverse plane. We have also introduced the quantities
\begin{eqnarray}
B_{RV} &=& 1 - \frac{2 \pi^2}{3} \epsilon^2-\frac{14}{3} \zeta_3 \epsilon^3+\frac{\pi^4}{15} \epsilon^4, \nonumber \\
A_{ij,k}(x,\phi) &=& \frac{n_i \cdot n_k}{n_i \cdot n_j} + x \frac{n_j \cdot n_k}{n_i \cdot n_j} - 2\,\text{cos}(\phi) 
	\frac{\sqrt{x \,n_i \cdot n_k\, n_j \cdot n_k}}{n_i \cdot n_j}.
\end{eqnarray}
Poles in $\epsilon$ occur in three places in $I^{(2),ij}_{RV,i}$: from the explicit overall factor of $1/\epsilon^2$; from the term $\taun^{-1-4\epsilon}$ in the limit $\taun \to 0$; from the limit $s \to 0$ of the term $s^{-1+2\epsilon}$ in the integrand.  The presence of the theta function makes this expression difficult to integrate analytically and extract the pole in $s$.  However, the poles can be easily extracted using plus-distribution expansions in $s$ and $\taun$:
\begin{equation} \label{eq:plusexp}
x^{-1+\epsilon} = \frac{1}{\epsilon} \delta(x) +\sum_{n=0} \frac{\epsilon^n}{n!} \left[ \frac{\text{ln}^n \,x}{x} \right]_+,
\end{equation}
where $x$ denotes either $s$ or $\taun$.  One this is done the resulting coefficients of the Laurent expansion in $\epsilon$ can be easily integrated numerically.  The case $r=j$ is easily obtained by permuting the indices $i$ and $j$ in the quantity $A_{ij,k}$.  We note that the $s \to 0$ limit is associated with an ultraviolet singularity, since $q^-  \to \infty$.  In SCET$_{\text I}$ such singularities can always be regulated in dimensional regularization, mapped to the unit hypercube and extracted with a variable change of the type used here.

The second case with $r=k$ proceeds similarly, except that the Sudakov decomposition of the radiation momentum instead uses $n_m = n_k$, $n_n = n_i$.  We denote this contribution with the subscript $k$.  Proceeding as before, we derive the final result 
\begin{equation}
\begin{split}
I^{(2),ij}_{RV,k}(\taun) &= -\frac{C_A}{\epsilon^2} B_{RV} \, \taun^{-1-4\epsilon} ( n_i \cdot n_j)^{1+\epsilon} \, ( n_i \cdot n_k)^{-1+\epsilon} \, 2^{-1-4\epsilon} 	\int_0^1 ds \, dx_2 \, s^{3\epsilon} \,  \text{sin}^{-2\epsilon}(\phi_1) \\ &\times \left[ A_{ki,j}(s,\phi_1) \right]^{-1-\epsilon}\,
	\prod_{l \neq i,k} \theta\left[ A_{ki,l}(s,\phi_{1l}) -s \right].
\end{split}
\end{equation}
This form is again suitable for numerical implementation, as the quantity $A_{ki,j}$ remains finite throughout the allowed phase space.  Using $I^{(2),ij}_{RV,i}$ and $I^{(2),ij}_{RV,k}$, the entire real-virtual contribution to the NNLO soft function for $N$-jettiness can be derived.

\subsection{The $q\bar{q}$ double-real correction} \label{subsec:qqb}

\begin{figure}[!h]
\begin{center}
\includegraphics[width=0.55\textwidth,angle=0]{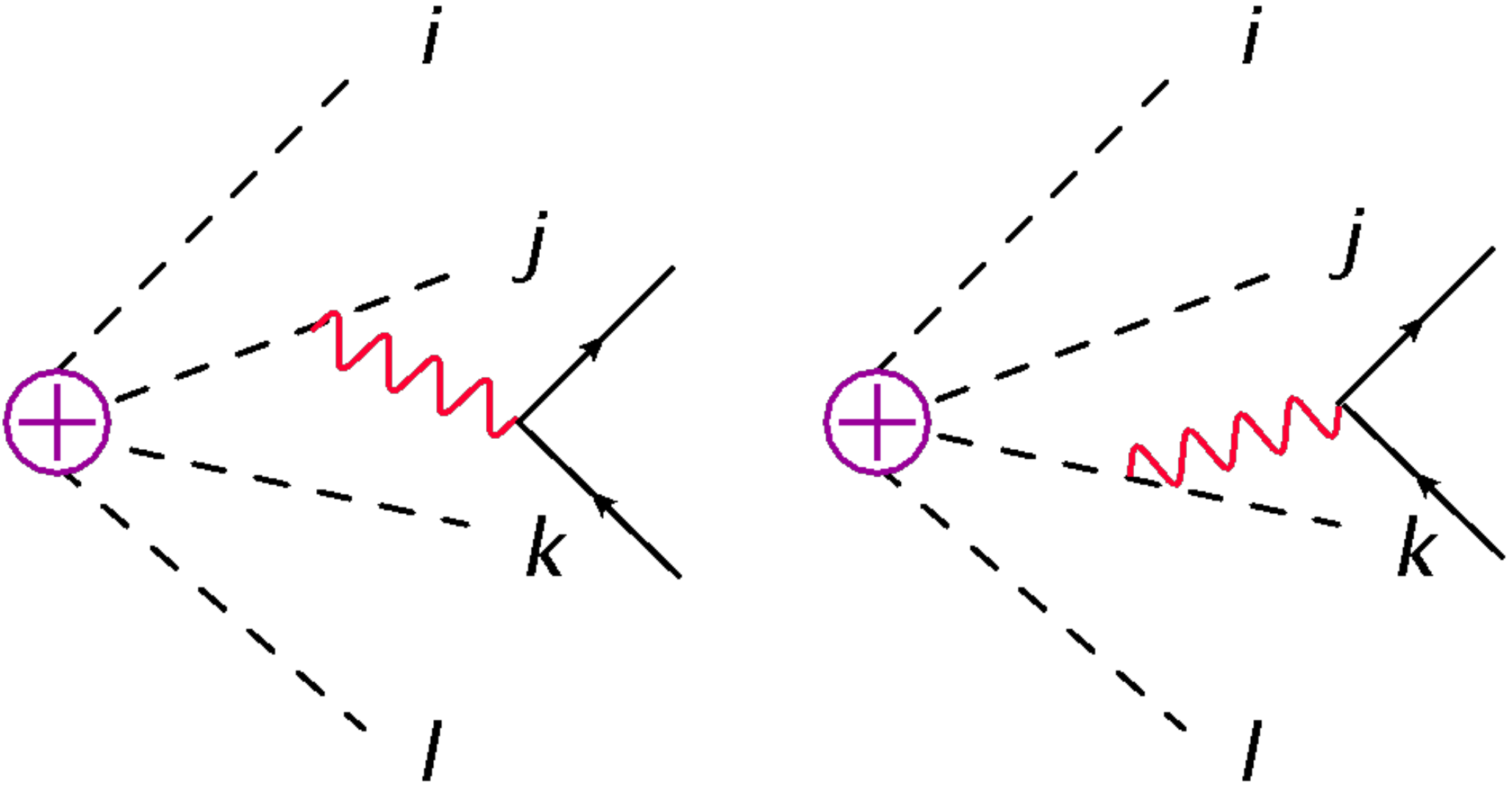}
\end{center}
\vspace{-0.5cm}
\caption{Representative diagrams arising from emission of a $q\bar{q}$ pair.  The dashed lines represent the eikonal directions.} \label{fig:RRqqb}
\end{figure}

We next consider the correction arising from the emission of a $q\bar{q}$ pair from the hard eikonal lines.  Several representative diagrams are shown in Fig.~\ref{fig:RRqqb}.  The integrand can again be obtained from the QCD result for the emission of a soft $q\bar{q}$ pair, as presented in Ref.~\cite{Catani:1999ss}.  It is expressed in terms of the function 
\begin{equation}
{\cal I}_{ij}(q_1,q_2) = \frac{\dotp{p_i}{q_1}\dotp{p_j}{q_2}+\dotp{p_j}{q_1}\dotp{p_}{q_2}-\dotp{p_i}{q_j}\dotp{q_1}{q_2}}
	{(\dotp{q_1}{q_2})^2 [\dotp{p_i}{(q_1+q_2)}] [\dotp{p_j}{(q_1+q_2)}]}.
\end{equation}
However, color conservation of QCD amplitudes restricts the ways in which ${\cal I}_{ij}$ enters the soft function.  The factorization of the QCD amplitude in the double-soft limit indicates that
\begin{equation} \label{eq:QCDfac}
|{\cal M}(\ldots,q_1,q_2)|^2 \approx \langle {\cal M} | \left( \sum_{i,j} {\cal I}_{ij}\, \bold{T}_i \cdot \bold{T}_j  \right)| {\cal M}\rangle,
\end{equation}
where we have used color-space notation~\cite{Catani:1996vz} in writing the color-correlated product of amplitudes on the right-hand side.  Color conservation allows us to write
\begin{equation}\label{eq:colcons}
\sum_j \bold{T_j} | {\cal M}\rangle= 0.
\end{equation}
We use this relation to remove all color structures of the form $\bold{T}_i \cdot \bold{T}_i$ in Eq.~(\ref{eq:QCDfac}), by dotting Eq.~(\ref{eq:colcons}) with $\bold{T}_i$ and solving for the $\bold{T}_i \cdot \bold{T}_i$ term.  Doing so, we find that the coefficient of each remaining color structure $T_i \cdot T_j$, with $i \neq j$, contains the combination
\begin{equation}
{\cal J}_{ij} = {\cal I}_{ii}+{\cal I}_{jj}-2 {\cal I}_{ij}.
\end{equation}
This is the basic building block of the $q\bar{q}$ contribution whose integration over phase space we will study.

The integral we must consider is
\begin{equation}
I^{(2),ij}_{q\bar{q}}(\taun) = 64\pi^4 e^{2\epsilon\gamma_E} (4\pi)^{-2\epsilon} \left( \frac{\alpha_s}{2\pi}\right)^2
	 \int \frac{d^{d-1}q_1}{(2\pi)^{d-1}}  \frac{d^{d-1}q_2}{(2\pi)^{d-1}} \, {\cal J}_{ij}(q_1,q_2)\, {\cal M}(q_1,q_2).
\end{equation}
The overall numerical factor comes from expressing the bare coupling constant in terms of the renoramlized one.  We have kept explicit the overall coefficient of
$(\alpha_s/(2\pi))^2$ in order to make clear the normalization of our result.  It is convenient to divide this integral into several different structures, according to which phase-space parameterization is most suitable for performing 
the extraction of singularities.  We first divide it into regions according to whether the radiated quarks $q_1$ and $q_2$ are closest to one of the emitting eikonal lines $i,j$, or are closer to non-emitting lines which we label as $k,l$.  This leads us to five distinct regions to investigate:
\begin{enumerate}

\item both $q_1$ and $q_2$ closest to the same emitting direction $i$, which we denote as $I^{(2),ij}_{q\bar{q},ii}$;
\item $q_1$ and $q_2$ closest to different emitting directions, which we denote as $I^{(2),ij}_{q\bar{q},ij}$;
\item $q_1$ closest to $i$ and $q_2$ nearest to $k$, which we call $I^{(2),ij}_{q\bar{q},ik}$;
\item both $q_1$ and $q_2$ closest to $k$, which we call $I^{(2),ij}_{q\bar{q},kk}$;
\item $q_1$ and $q_2$ closest to different non-emitting directions, which we denote as $I^{(2),ij}_{q\bar{q},kl}$.

\end{enumerate}
We furthermore find it convenient to divide the integrand ${\cal J}_{ij}$ into two structures according to whether the denominator is quadratic or linear in the invariant $q_1 \cdot q_2$.  We label these as $I$ and $II$, respectively, so that ${\cal J}_{ij} = {\cal J}_{ij}^{I}+{\cal J}_{ij}^{II}$.  Written explicitly, the integrands for these two structures are
\begin{eqnarray}
{\cal J}_{ij}^{I} &=& -2 \frac{[\dotp{p_i}{q_1}\dotp{p_j}{q_2}+\dotp{p_j}{q_1}\dotp{p_}{q_2}]^2}{(\dotp{q_1}{q_2})^2 [\dotp{p_i}{(q_1+q_2)}]^2 [\dotp{p_j}{(q_1+q_2)}]^2}, \nonumber \\
{\cal J}_{ij}^{II} &=& 2 \frac{\dotp{p_i}{p_j}}{(\dotp{q_1}{q_2}) [\dotp{p_i}{(q_1+q_2)}] [\dotp{p_j}{(q_1+q_2)}]}.
\end{eqnarray}
This leaves us with a total of ten integrals to compute.  The entire $q\bar{q}$ contribution to the soft function can be obtained by appropriately permuting the indices of these structures.  Since the computation proceeds similarly for all ten terms, we will focus on the representative example $I^{(2),ij,II}_{q\bar{q},kk}$ which exhibits all of the complexities that must be addressed in the general case.

We begin by performing a Sudakov decomposition of both $q_1$ and $q_2$ as in Eq.~(\ref{eq:Suddecomp}), choosing the light-cone directions $m=k$ and $n=i$.   The explicit representations of the transverse vectors $q_{1\perp}$ and $q_{2\perp}$ are as follows:
\begin{eqnarray} \label{eq:qperpparam}
q_{1\perp}^{\mu} &=& |q_{1\perp}| \left( \text{cos}(\phi_1),  \text{sin}(\phi_1); 0\right), \nonumber \\
q_{2\perp}^{\mu} &=& |q_{2\perp}| \left( \text{cos}(\phi_2),  \text{sin}(\phi_2) \, \text{cos}(\alpha); \text{sin}(\phi_2) \, \text{sin}(\alpha) \, \hat{n}_{\epsilon}\right).
\end{eqnarray}
The last component of each momentum appearing after the semi-colon denotes the $\epsilon$-dimensional component of transverse momentum (We recall that the dimensionality of the transverse plane is $2-2\epsilon$ in dimensional regularization).  $\hat{n}_{\epsilon}$ denotes a unit vector in the $-2\epsilon$-dimensional space.  A single angle is needed in the general $N$-jettiness case to parameterize this direction. It is now straightforward to write down the following expression for the integral:
\begin{equation}
\begin{split}
I^{(2),ij,II}_{q\bar{q},kk} &=  2^{1-8\epsilon}\, B_{RR} \left( \frac{\alpha_s}{2\pi}\right)^2 \taun^{-1-4\epsilon} \,  \dotp{n_i}{n_j} [\dotp{n_i}{n_k}]^{-1+2\epsilon}
	\int_0^1 d\xi \,ds\, dt\, dx_4\,dx_5\,dx_6 \, d\Omega^{(\epsilon)} \\ & \times[\xi(1-\xi)]^{-2\epsilon}   [\lambda(1-\lambda)]^{-\epsilon}\text{sin}^{-2\epsilon}(\phi_1) 
	[-\epsilon x_6^{-1-\epsilon}] (1-x_6)^{-\epsilon} [s\,t]^{1+\epsilon} |s-t|^{-1-2\epsilon} \\ 
	&\times \frac{\left\{ (\sqrt{s}-\sqrt{t})^2+4 \lambda \sqrt{st} \right\}^{2\epsilon}}{\xi \, t A_{ki,j}(s,\phi_1)+(1-\xi)sA_{ki,j}(t,\phi_{2k})}\frac{1}{\xi t+(1-\xi)s}
	\\& \times \prod_{l \neq i,k} \theta\left[ A_{ki,l}(s,\phi_{1l}) -s \right] \theta\left[ A_{ki,l}(t,\phi_{2l}) -t \right],
\end{split}
\end{equation}
where we have introduced the abbreviation
\begin{equation}
B_{RR} = 1 - \frac{\pi^2}{3} \epsilon^2-\frac{8}{3} \zeta_3 \epsilon^3+\frac{\pi^4}{90} \epsilon^4.
\end{equation}
We have made the following variable changes to arrive at this expression:
\begin{equation}
\begin{split}
q_1^+  &= \taun \xi,\;\;\; q_1^- = \frac{\taun \xi}{s},\;\;\; q_2^+  = \taun (1-\xi),\;\;\; q_2^- = \frac{\taun (1-\xi)}{t}, \\
\phi_1 &= 2\pi x_4, \;\;\; \lambda = \text{sin}^2(\pi x_5/2), \;\;\; \text{cos}(\alpha) = 1-2 x_6.
\end{split}
\end{equation}
We have in addition followed the sector decomposition approach to NNLO calculations~\cite{Anastasiou:2003gr,Czakon:2010td,Boughezal:2011jf} and have made a non-linear change of variables to map $\text{cos}(\phi_2)$ to the unit hypercube.  The quantity $d\Omega^{(\epsilon)}$ denotes the angular parameterization of the direction $\hat{n}_{\epsilon}$ in Eq.~(\ref{eq:qperpparam}), normalized so that it integrates to unity.  We note that in the zero-jettiness and one-jettiness cases, we may immediately integrate over 
$d\Omega^{(\epsilon)}$ to obtain unity.

This integral is not yet suitable for numerical implementation, as there are singularities associated with the joint limit $s,t \to 0$ that cannot yet be extracted with a plus-distribution expansion.  We order these two limits by inserting the following partition into phase space, as is done in sector decomposition of real radiation in QCD~\cite{Anastasiou:2003gr,Czakon:2010td,Boughezal:2011jf}:
\begin{equation}
1 = \theta(s-t)+\theta(t-s).
\end{equation}
We then remap the limits of integration to the unit hypercube.  Doing so renders all singularities amenable to a plus distribution example.  Focusing on the $s>t$ sector for illustrative purposes and setting $\xi = x_1$, $s=x_2$, and $t = x_2(1-x_3)$, we find the following final result:
\begin{equation}
\begin{split}
I^{(2),ij,II}_{q\bar{q},kk,s>t} &=  2^{1-8\epsilon}\, B_{RR} \left( \frac{\alpha_s}{2\pi}\right)^2 \taun^{-1-4\epsilon} \,  \dotp{n_i}{n_j} [\dotp{n_i}{n_k}]^{-1+2\epsilon}
	\int_0^1 dx_1 \,dx_2\, dx_3\, dx_4\,dx_5\,dx_6\, d\Omega^{(\epsilon)} \\ & \times[x_1(1-x_1)]^{-2\epsilon}   [\lambda(1-\lambda)]^{-\epsilon}\text{sin}^{-2\epsilon}(\phi_1) 
	[-\epsilon x_6^{-1-\epsilon}] (1-x_6)^{-\epsilon} x_2^{2\epsilon}(1-x_3)^{1+\epsilon} x_3^{-1-2\epsilon}\\ 
	&\times \frac{\left\{ (1-\sqrt{1-x_3})^2+4 \lambda \sqrt{1-x_3} \right\}^{2\epsilon}}{x_1(1-x_3) \, t A_{ki,j}(x_2,\phi_1)
		+(1-x_1) A_{ki,j}(x2(1-x_3),\phi_{2k})}\frac{1}{1-x_1x_3}
	\\& \times \prod_{l \neq i,k} \theta\left[ A_{ki,l}(x_2,\phi_{1l}) -x_2 \right] \theta\left[ A_{ki,l}(x_2(1-x_3),\phi_{2l}) -x_2(1-x_3) \right].
\end{split}
\end{equation}
The poles in this expression come only from expanding the overall $\taun^{-1-4\epsilon}$ and the factor $x_3^{-1-2\epsilon}$ using the plus distribution expansion of Eq.~(\ref{eq:plusexp}).  Once this expansion in performed, the integral can be simply evaluating numerically.  The computations of the other sector and the other integrals required for the $q\bar{q}$ contribution proceed similarly.  Only the ordering of the $s$ and $t$ limits is needed to make all singularities manifest.

\subsection{The $gg$ double-real correction}

\begin{figure}[!h]
\begin{center}
\includegraphics[width=0.55\textwidth,angle=0]{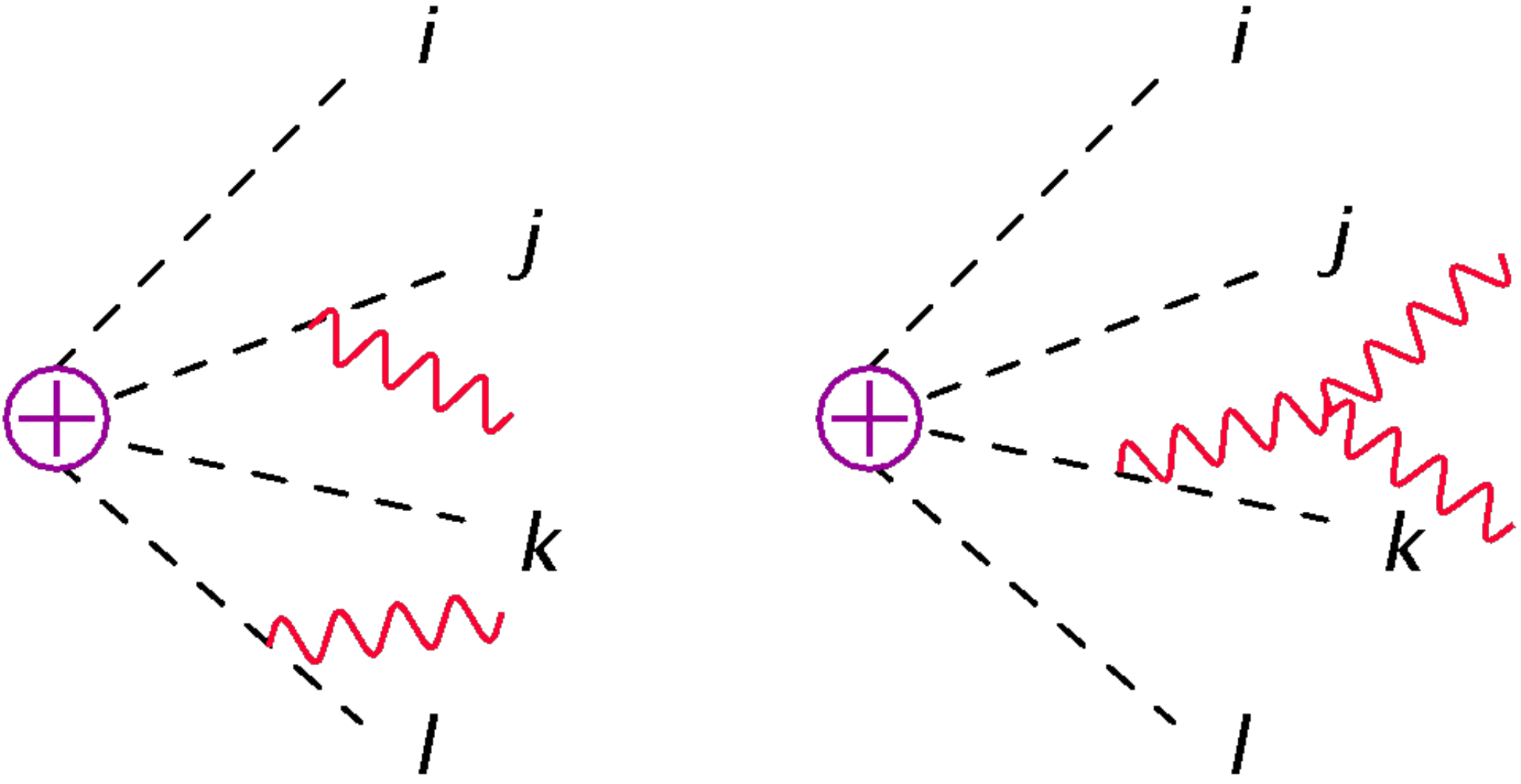}
\end{center}
\vspace{-0.5cm}
\caption{Representative diagrams arising from emission of a $gg$ pair.  The dashed lines represent the eikonal directions.} \label{fig:RRgg}
\end{figure}

Finally, we consider the correction arising from the emission of two gluons from the hard eikonal lines.  Several representative diagrams are shown in Fig.~\ref{fig:RRgg}.   The integrand can be obtained from Ref.~\cite{Catani:1999ss}.  Two distinct types of contributions to the soft function can be identified.  The first type comes from the squares of single-gluon currents, and can be easily obtained following the techniques of Ref.~\cite{Jouttenus:2011wh}.  We do not discuss it further here.  The second type are genuinely non-abelian contributions proportional to $C_A$ that are not simply the square of one-loop terms.  These can be written in the form of Eq.~(\ref{eq:QCDfac}), except with the replacement of eikonal functions ${\cal I}_{ij} \to {\cal S}_{ij}$, with $ {\cal S}_{ij}$ given in Eq.~(110) of Ref.~\cite{Catani:1999ss}.  The same color conservation argument as for the $q\bar{q}$ case indicates that only the combination
\begin{equation}
{\cal T}_{ij} = {\cal S}_{ii}+{\cal S}_{jj}-2 {\cal S}_{ij}
\end{equation}
contributes to the result.  Simple algebraic manipulation leads us to the result
\begin{eqnarray} \label{eq:integrandgg}
{\cal T}_{ij} &=& (1-\epsilon) {\cal J}_{ij}^{I}+2 {\cal J}_{ij}^{II} +\left(\frac{\dotp{p_i}{q_1}\dotp{p_j}{q_2}
	+\dotp{p_j}{q_1}\dotp{p_i}{q_2}}{[\dotp{p_i}{(q_1+q_2)}] [\dotp{p_j}{(q_1+q_2)}]} -2\right) \, S_{ij}^{(s.o.)}, \nonumber \\
S_{ij}^{(s.o.)} &=& \frac{\dotp{p_i}{p_j}}{\dotp{q_1}{q_2}} \left(\frac{1}{\dotp{p_i}{q_1}\dotp{p_j}{q_2}}+\frac{1}{\dotp{p_j}{q_1}\dotp{p_i}{q_2}} \right)
	-\frac{(\dotp{p_i}{p_j})^2}{\dotp{p_i}{q_1}\dotp{p_i}{q_2}\dotp{p_j}{q_1}\dotp{p_j}{q_2}}.		
\end{eqnarray}
The first two terms with ${\cal J}_{ij}^{I}$ and ${\cal J}_{ij}^{II}$ are the same structures as found in the $q\bar{q}$ case.  The structure proportional to the function $S_{ij}^{(s.o.)}$ arises from the strongly-ordered limit of QCD, in which there is a hierarchy between the energies of the radiated gluons.  It is  straightforward to follow the same parameterizations and steps presented in Sec.~\ref{subsec:qqb} to render this structure suitable for numerical evaluation, and we do not repeat the details here.

\section{Numerical results for one-jettiness} \label{sec:numerics}

We present in this section numerical results for the NNLO contributions to $N$-jettiness, including validation against known results in the literature.  We focus on two example cases in order to illustrate our results.  We begin with one-jettiness in electron-proton collisions, which has received recent interest in the contexts of probing nuclear dynamics in electron-nucleus collisions~\cite{Kang:2013wca,Kang:2013lga}, and of improving jet phenomenology in deep inelastic scattering~\cite{Kang:2013nha,Kang:2014qba}.  We also consider one-jettiness in proton-proton collisions, for which the NNLO soft function is a necessary component of a recently introduced subtraction scheme for NNLO fixed-order calculations~\cite{wjet,hjet}.  The restriction to one-jettiness only simplifies the color structure of the integrands.  The basic building blocks are those presented in the previous section. 

\subsection{One-jettiness in $ep$ collisions}

\begin{figure}[!h]
\begin{center}
\begin{minipage}[b]{0.46\linewidth}
\includegraphics[width=\textwidth,angle=0]{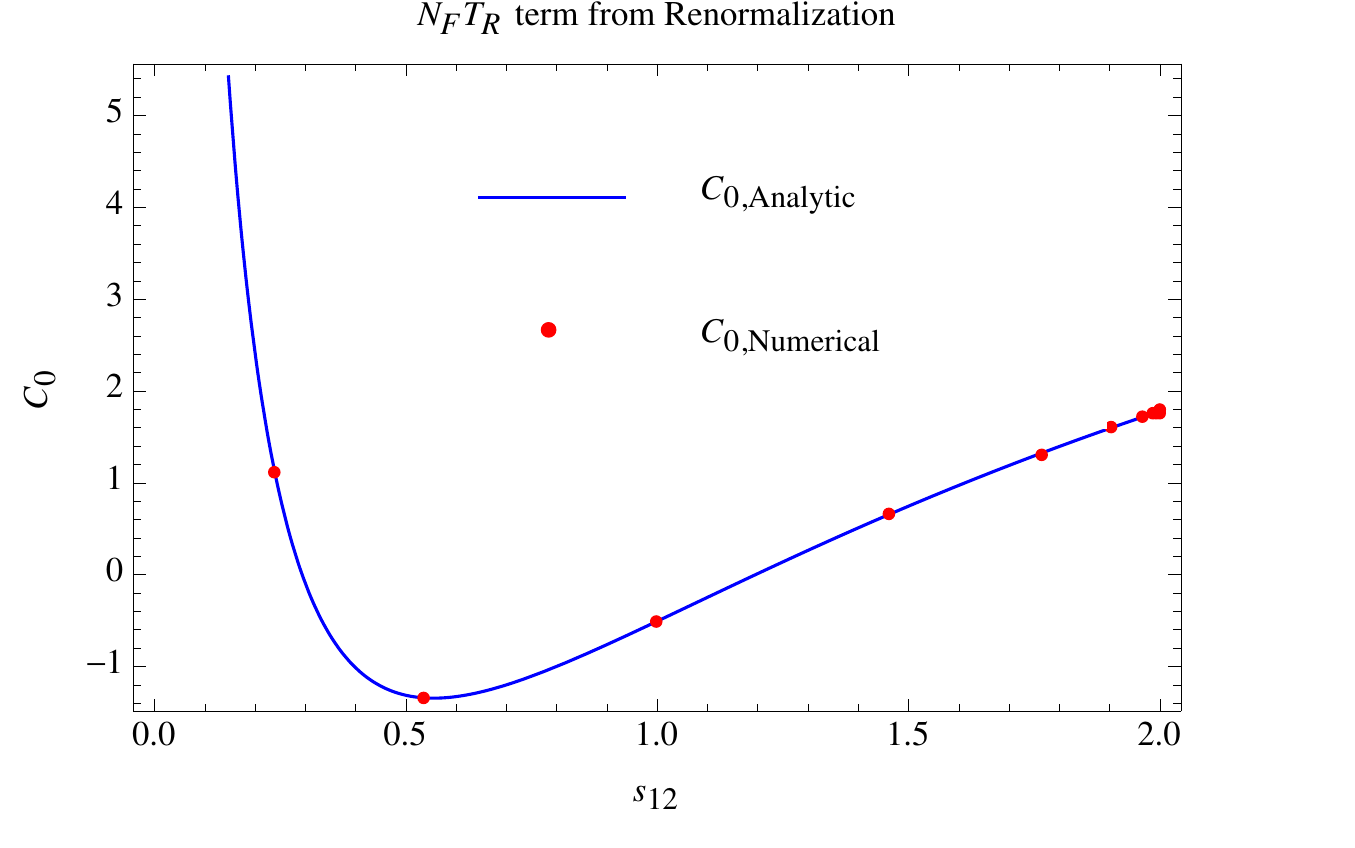}
\end{minipage}
\hspace{0.5cm}
\begin{minipage}[b]{0.46\linewidth}
\includegraphics[width=1.01\textwidth,angle=0]{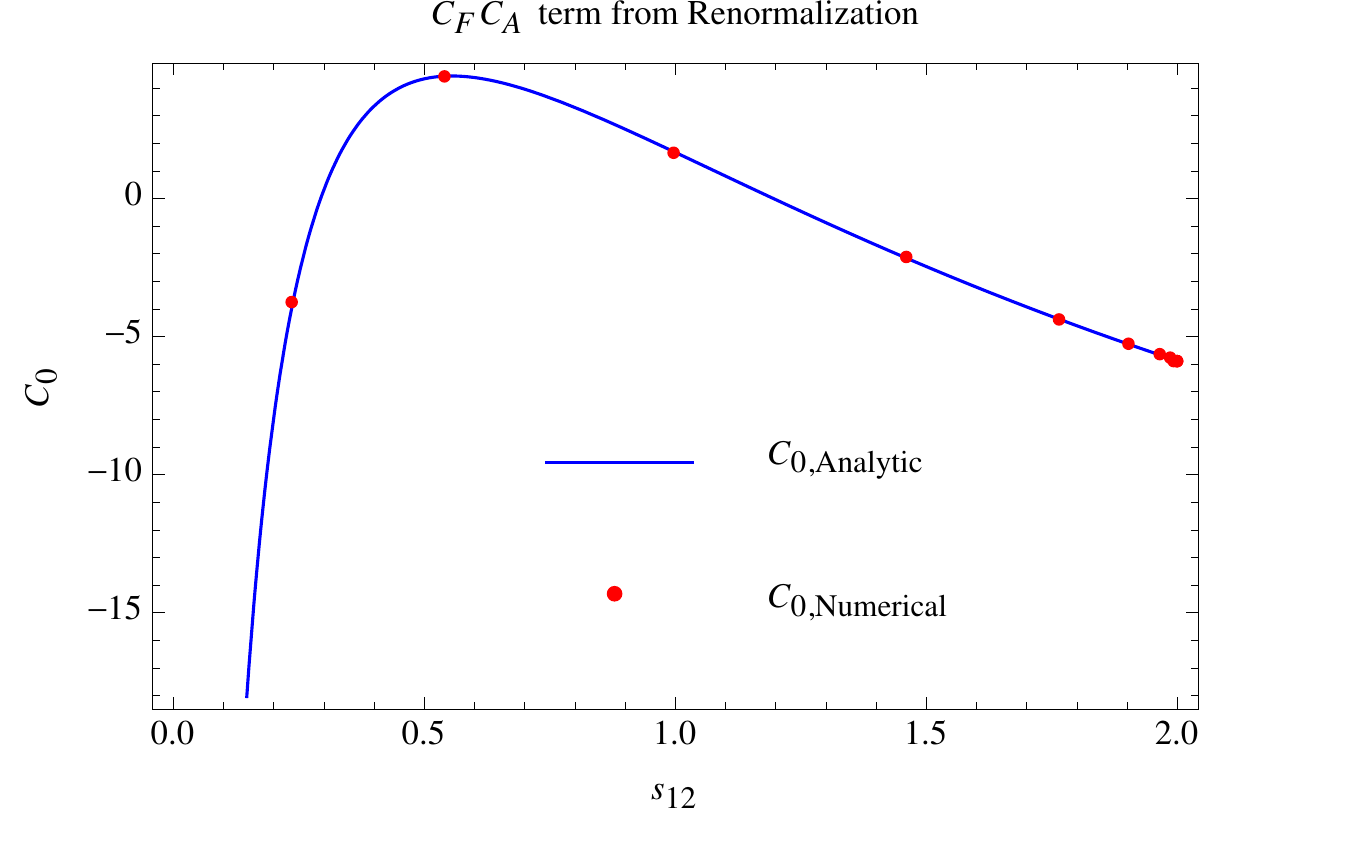}
\end{minipage}
\end{center}
\vspace{-0.5cm}
\caption{Comparison between the analytic and numerical calculations of the $C_0$ coefficients from renormalization, for both the
$N_F\, T_R$ (left panel) and $C_F\, C_A$ (right panel) terms, as a function of $s_{12}$. The blue solid lines represent the analytic calculation and the red dots are from our numerical approach. } \label{fig:ep-ren} 
\end{figure}

\begin{figure}[!h]
\begin{center}
\begin{minipage}[b]{0.46\linewidth}
\includegraphics[width=\textwidth,angle=0]{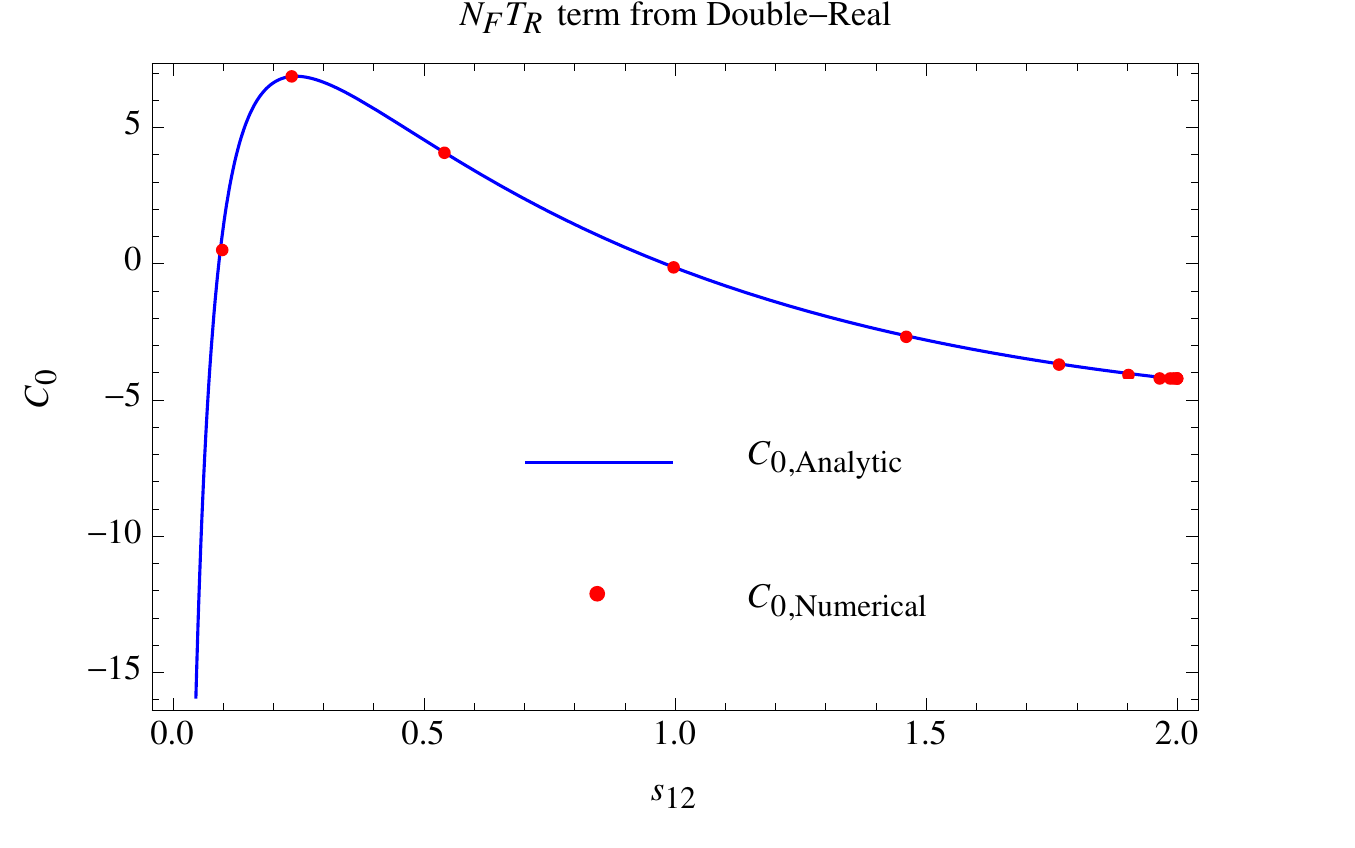}
\end{minipage}
\hspace{0.5cm}
\begin{minipage}[b]{0.46\linewidth}
\includegraphics[width=1.01\textwidth,angle=0]{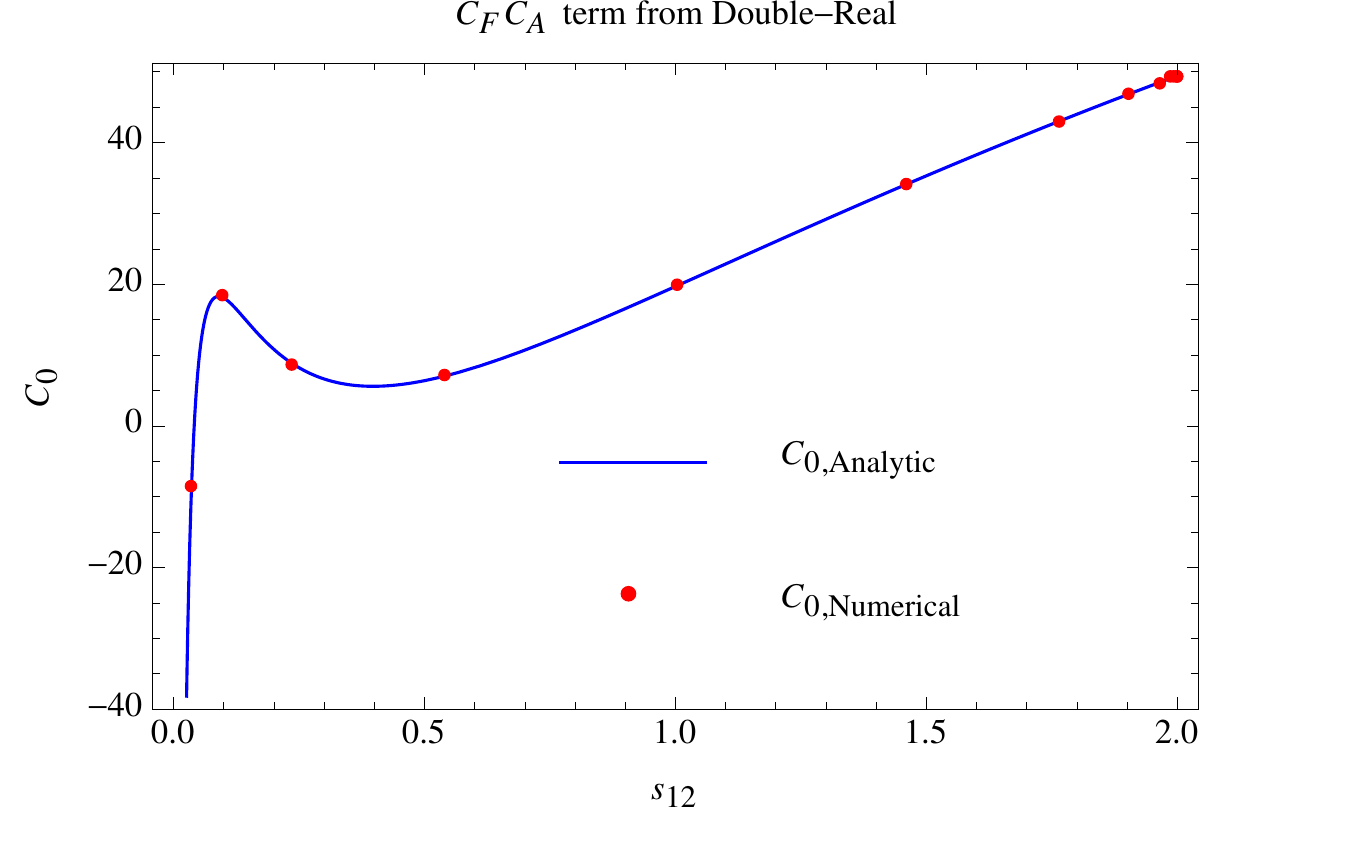}
\end{minipage}
\end{center}
\vspace{-0.5cm}
\caption{Comparison between the analytic and numerical calculations of the $C_0$ coefficients from the double-real contribution, for both the $N_F\, T_R$ (left panel) and $C_F\, C_A$ (right panel) terms, as a function of $s_{12}$.  The blue solid lines represent the analytic predictions and the red dots are from our numerical approach.} \label{fig:ep-rr}
\end{figure}

\begin{figure}[!h]
\begin{center}
\includegraphics[width=0.55\textwidth,angle=0]{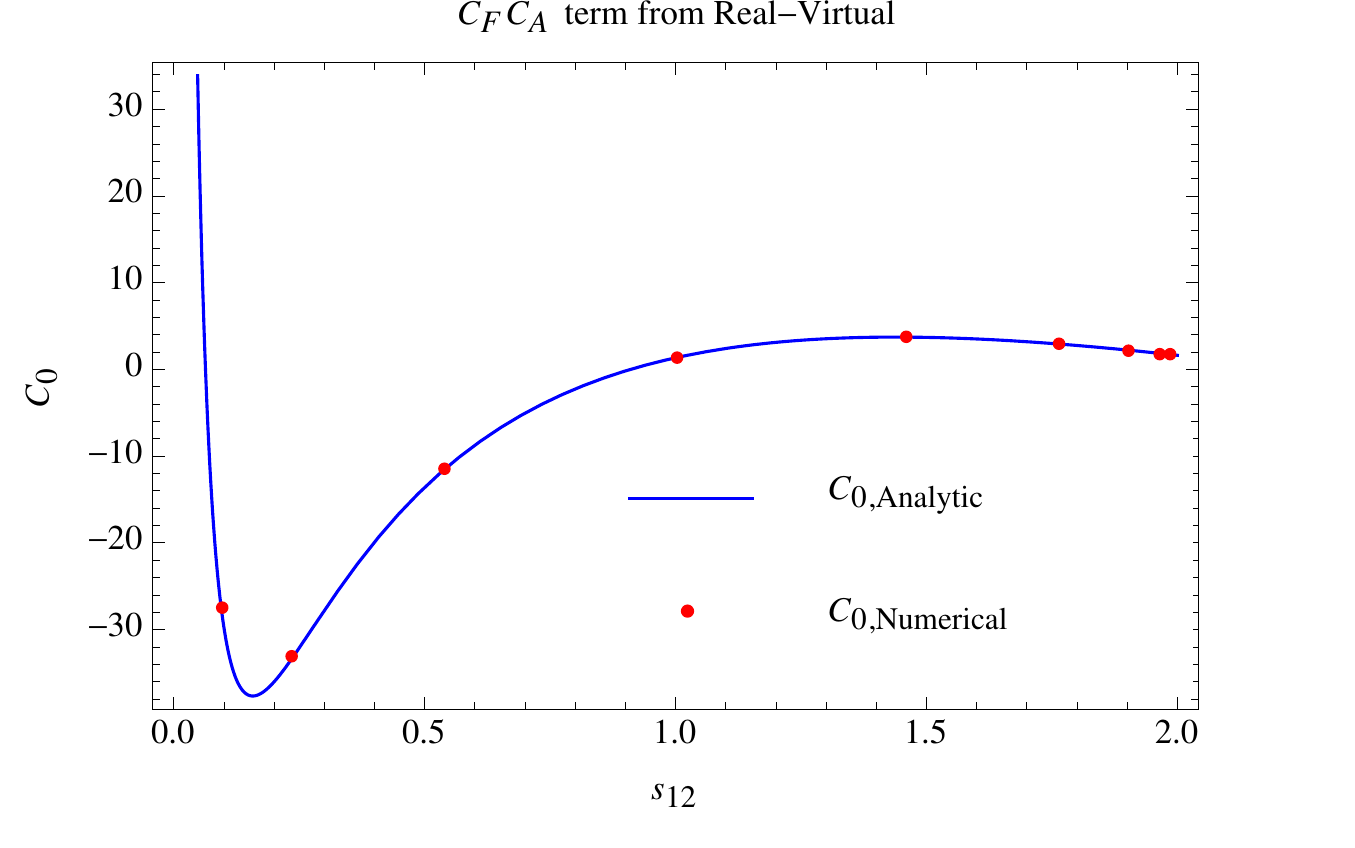}
\end{center}
\vspace{-0.5cm}
\caption{Comparison between the analytic and numerical calculations of the $C_0$ coefficients from the real-virtual contribution.  The blue solid lines represent the analytic predictions and the red dots are from our numerical approach.} \label{fig:ep-rv}
\end{figure}

We begin with a presentation of the one-jettiness soft function in $ep$ collisions as a validation of our calculation.  In this case the soft function can be obtained analytically from the known result for the thrust distribution in $e^+e^-$ collisions~\cite{Gehrmann:2012sc}.  We present the analytic result in the Appendix.  It is most convenient to express the result in terms of the soft contribution to the ${\cal O}(\alpha_s^2) $ cumulative cross section for one-jettiness:
\begin{equation}\label{cumxsec}
\Sigma_{soft}^{(2)}({\cal T}_1^{\rm cut } )  = \int_0^{{\cal T}_1^{\rm cut}} \,
\mathrm{d} {\cal T}_1 \frac{\mathrm{d} \sigma_{soft}^{(2)} }{\mathrm{d} {\cal T}_1 } 
= \left(\frac{\alpha_s}{2\pi} \right)^2 \left( 
C_4 \, L^4 + C_3 \, L^3 \, + C_2 \, L^2 \, + C_1\, L \, + C_0   
\right) \,,
\end{equation}
 with 
$L = \log ({\cal T}_1^{\rm cut } /\mu )$.  Since all components of the soft function contain the overall dependence ${\cal T}_1^{-1-4\epsilon}$, the integration over ${\cal T}_1$ to obtain this cumulant is trivial to perform.  We use the numerical approach described in the previous section and compare it against the analytic result for the $C_0$ coefficient.  We compare in Figs.~\ref{fig:ep-ren},~\ref{fig:ep-rr} and~\ref{fig:ep-rv} the separate contributions from the coupling-constant renormalization, the double-real emission contribution and the real-virtual correction.  In the former two cases, we further separate the $N_F\, T_R$ and $C_F\,C_A$ color structures.  Each contribution is plotted as a function of $s_{12} = n_1 \cdot n_2$.   In all cases, the numerical prediction for $C_0$ agrees perfectly with the analytic calculation.


\subsection{One-jettiness in $pp$ collisions}

We next consider the one-jettiness soft function in proton-proton collisions.  We calculate the ${\cal O}(\alpha_s^2)$ cumulative cross section defined in Eq.~(\ref{cumxsec}).  The one-jettiness soft function depends on three hard directions, which we label as $n_1$, $n_2$, and $n_3$.  We align $n_1$ and $n_2$ with the incoming beam axes in the $\pm z$ directions, and let $n_3$ lie along the outgoing jet direction.  The $C_i$ in 
Eq.~(\ref{cumxsec}) can then be written as functions of the kinematic invariant $s_{13} = n_1 \cdot n_3$. We note that $C_4$, $C_3$, $C_2$ and $C_1$ can be determined by expanding the resummed soft function to order $\alpha_s^2$. We use these terms to verify our direct fixed-order calculation.
We show the results for the Abelian and the non-Abelian contributions separately. In both cases we assume that the color factors associated with partons in the $n_1$, $n_2$ and $n_3$ directions are $C_F$, $C_A$ and $C_F$, respectively.
 
\begin{figure}[!h]
\begin{center}
\begin{minipage}[b]{0.46\linewidth}
\includegraphics[width=\textwidth,angle=0]{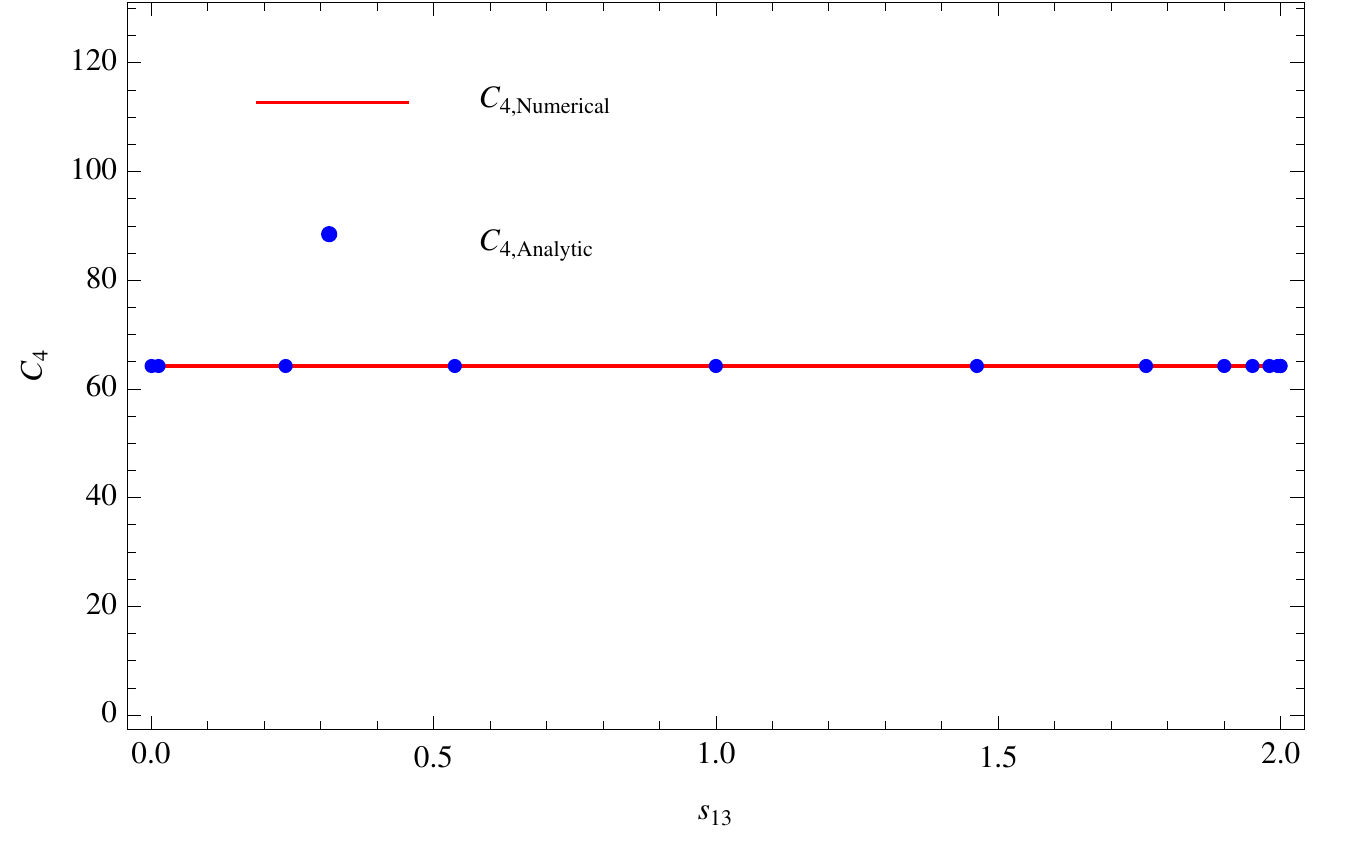}
\includegraphics[width=\textwidth,angle=0]{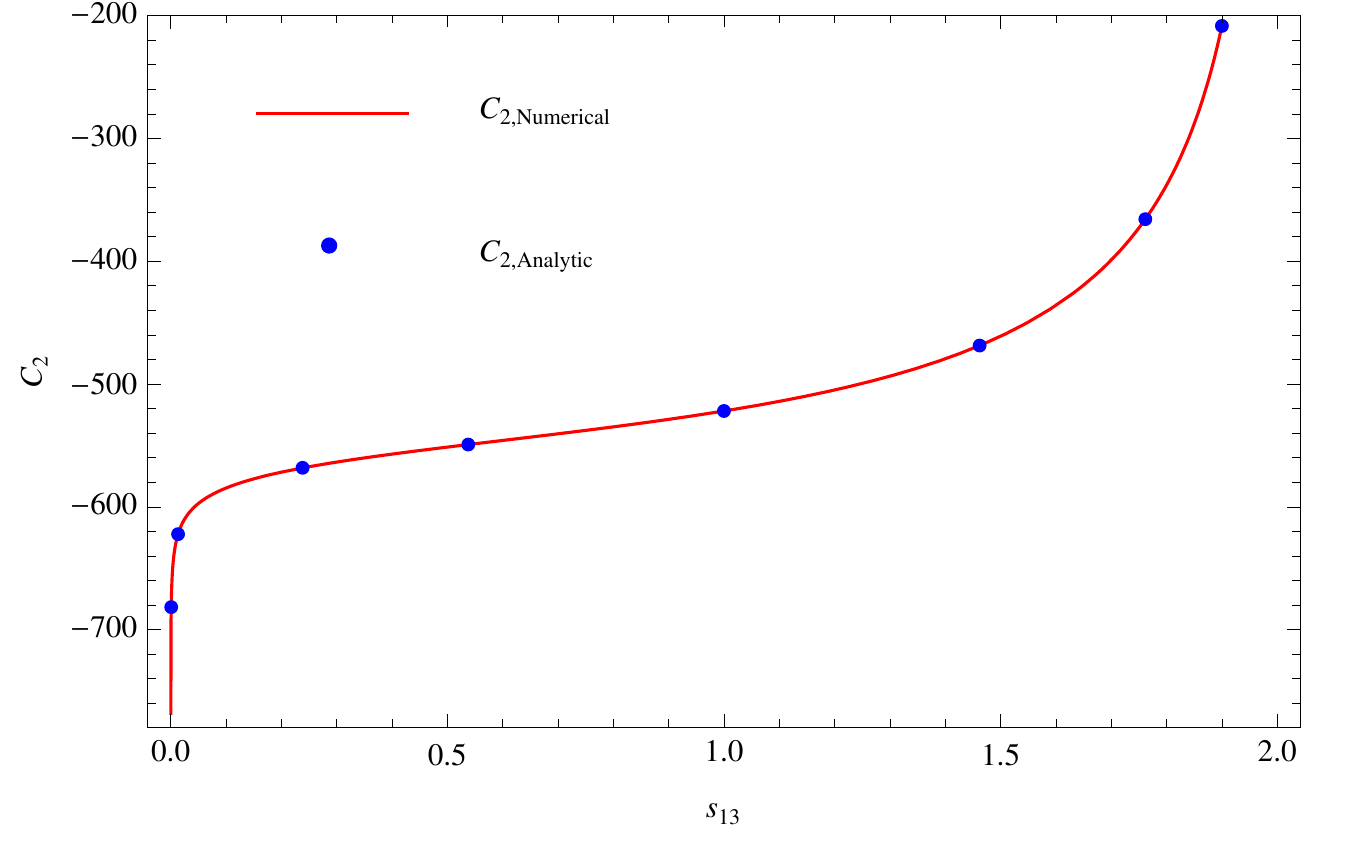}
\end{minipage}
\hspace{0.5cm}
\begin{minipage}[b]{0.46\linewidth}
\includegraphics[width=1.01\textwidth,angle=0]{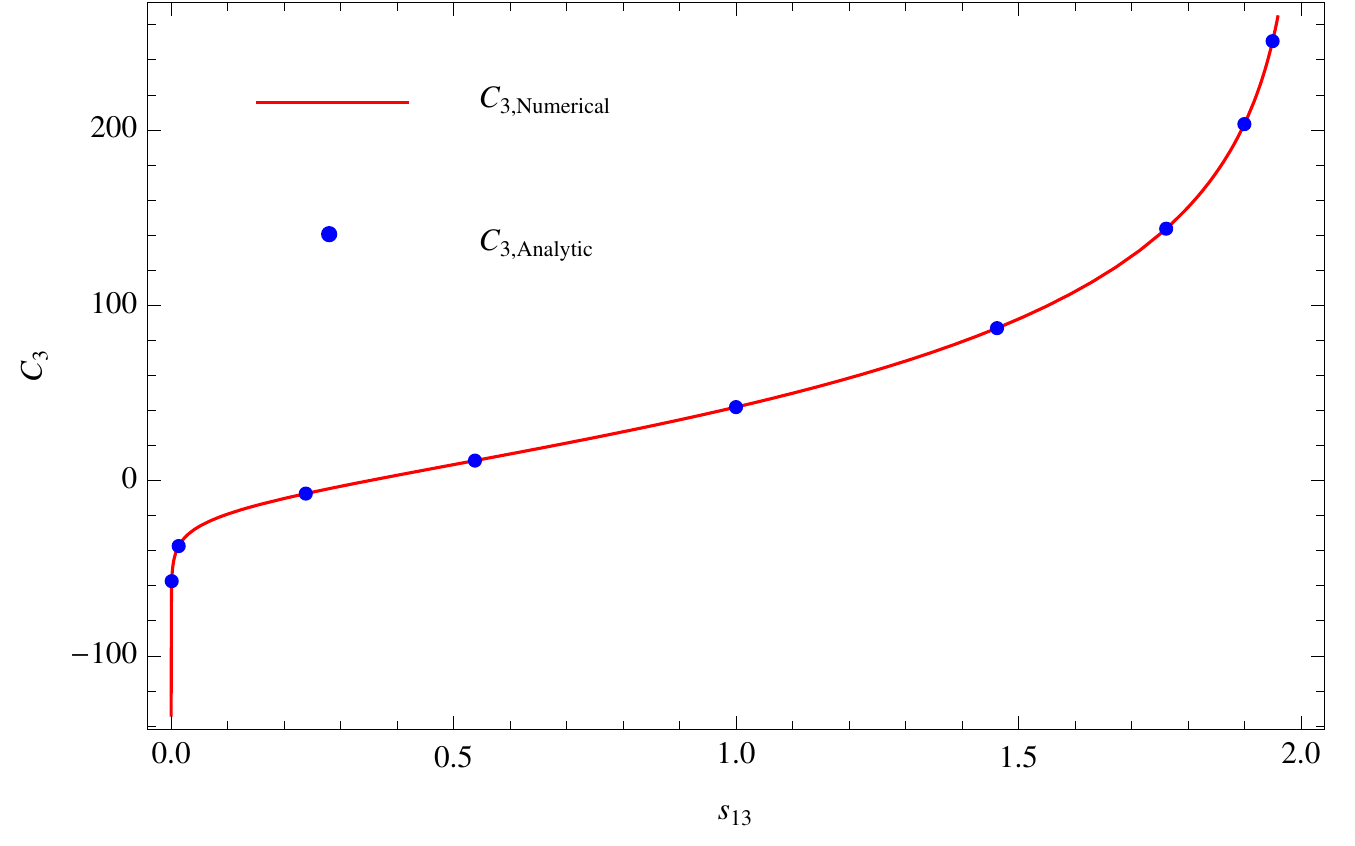}
\includegraphics[width=1.01\textwidth,angle=0]{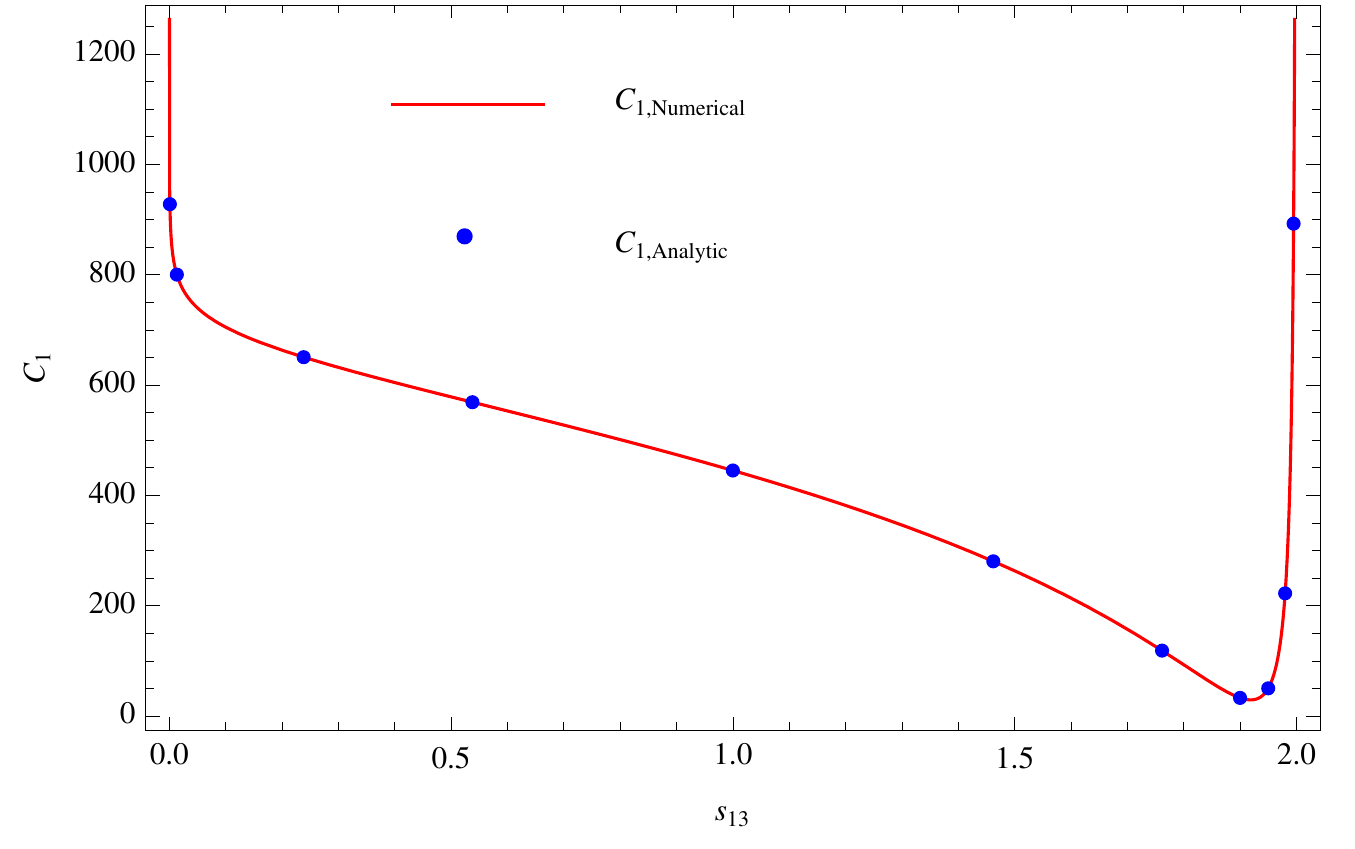}
\end{minipage}
\end{center}
\vspace{-0.5cm}
\caption{Comparison between the analytic and numerical calculation of the coefficients of $L^n$ ($n=4,3,2,1$) for the Abelian contribution to the soft function. The red solid lines represent our direct NNLO calculation, and the blue dots are predicted by expanding the resummed results to ${\cal O}(\alpha_s^2) $.} \label{fig:pp-abelian}
\end{figure}

\begin{figure}[!h]
\begin{center}
\begin{minipage}[b]{0.46\linewidth}
\includegraphics[width=\textwidth,angle=0]{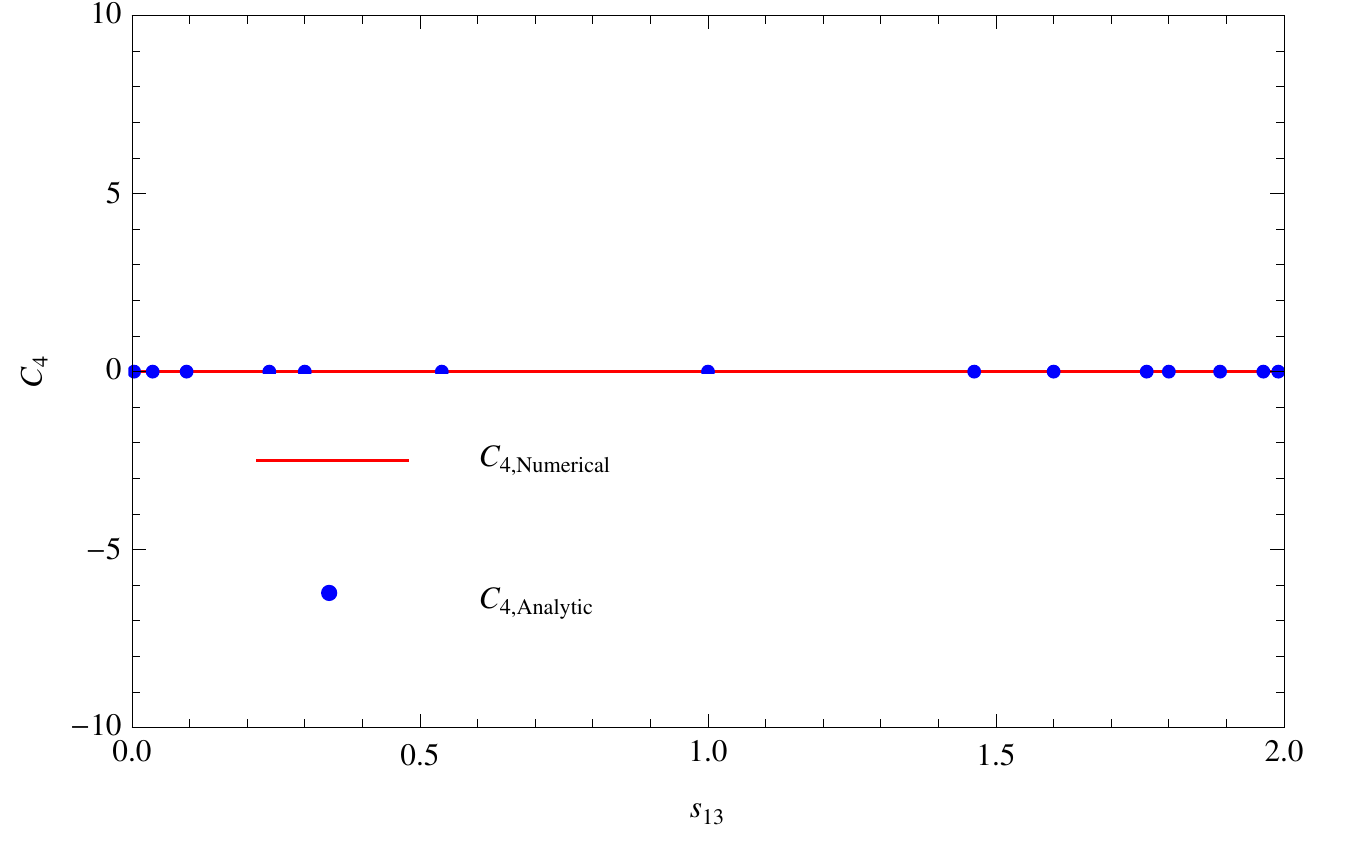}
\includegraphics[width=\textwidth,angle=0]{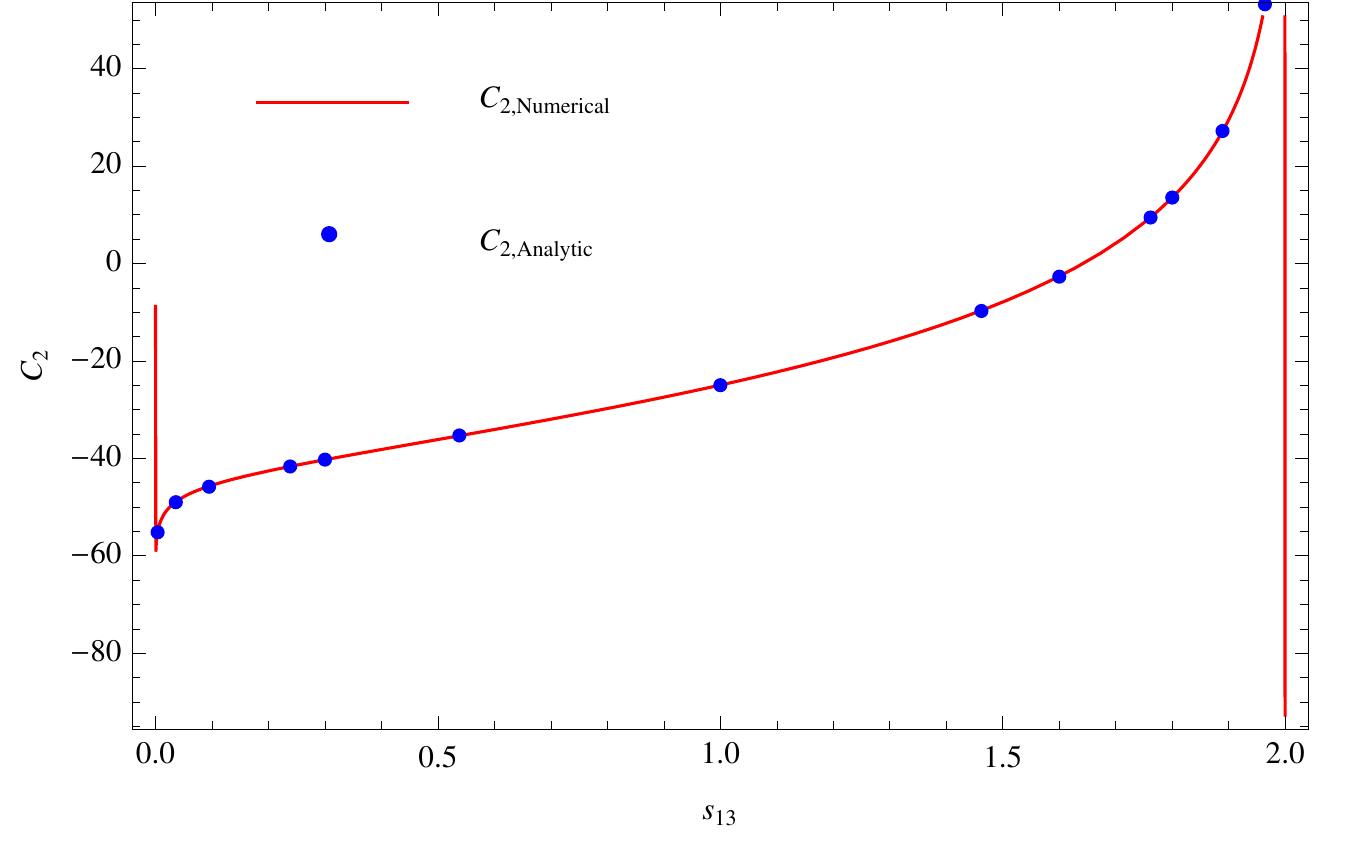}
\end{minipage}
\hspace{0.5cm}
\begin{minipage}[b]{0.46\linewidth}
\includegraphics[width=1.01\textwidth,angle=0]{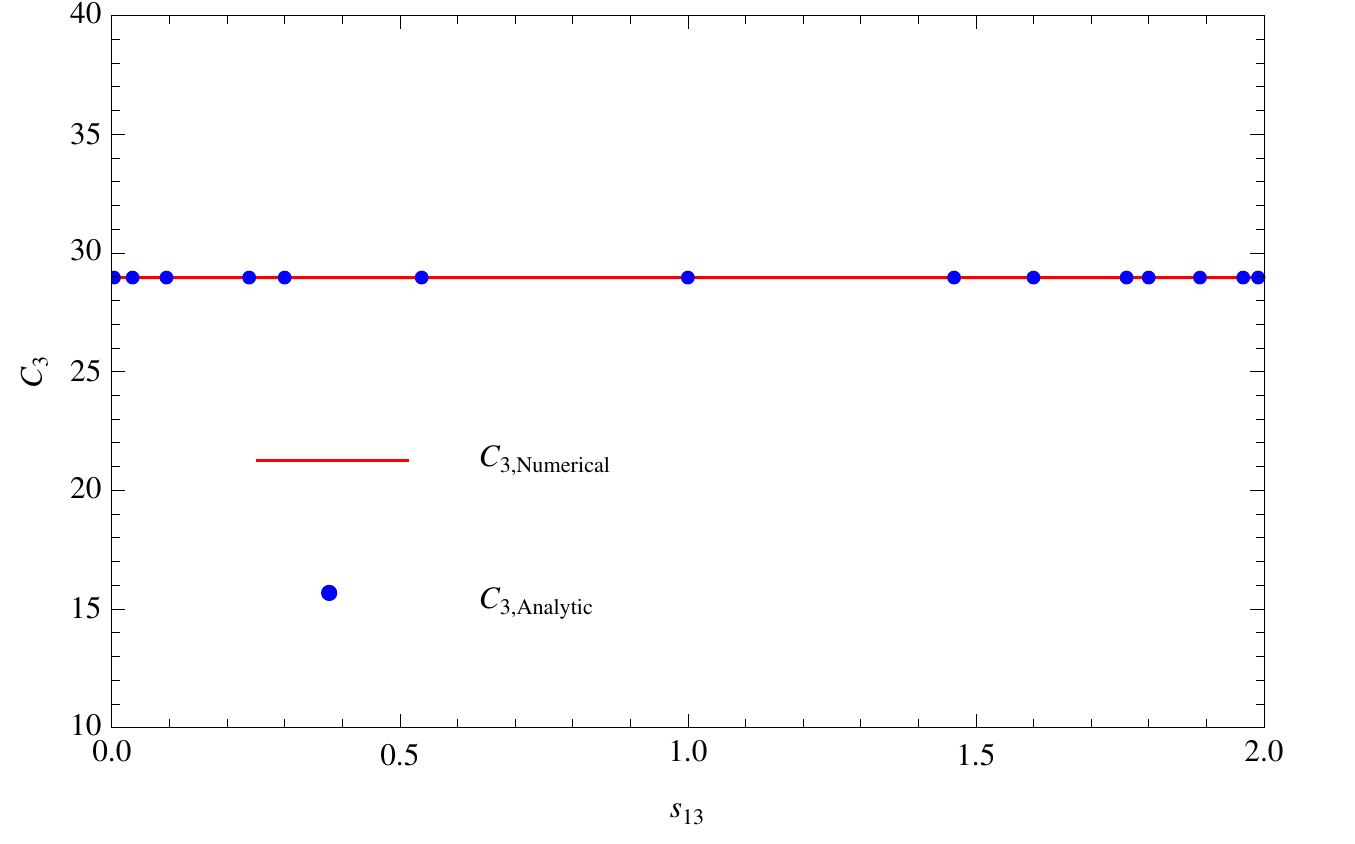}
\includegraphics[width=1.01\textwidth,angle=0]{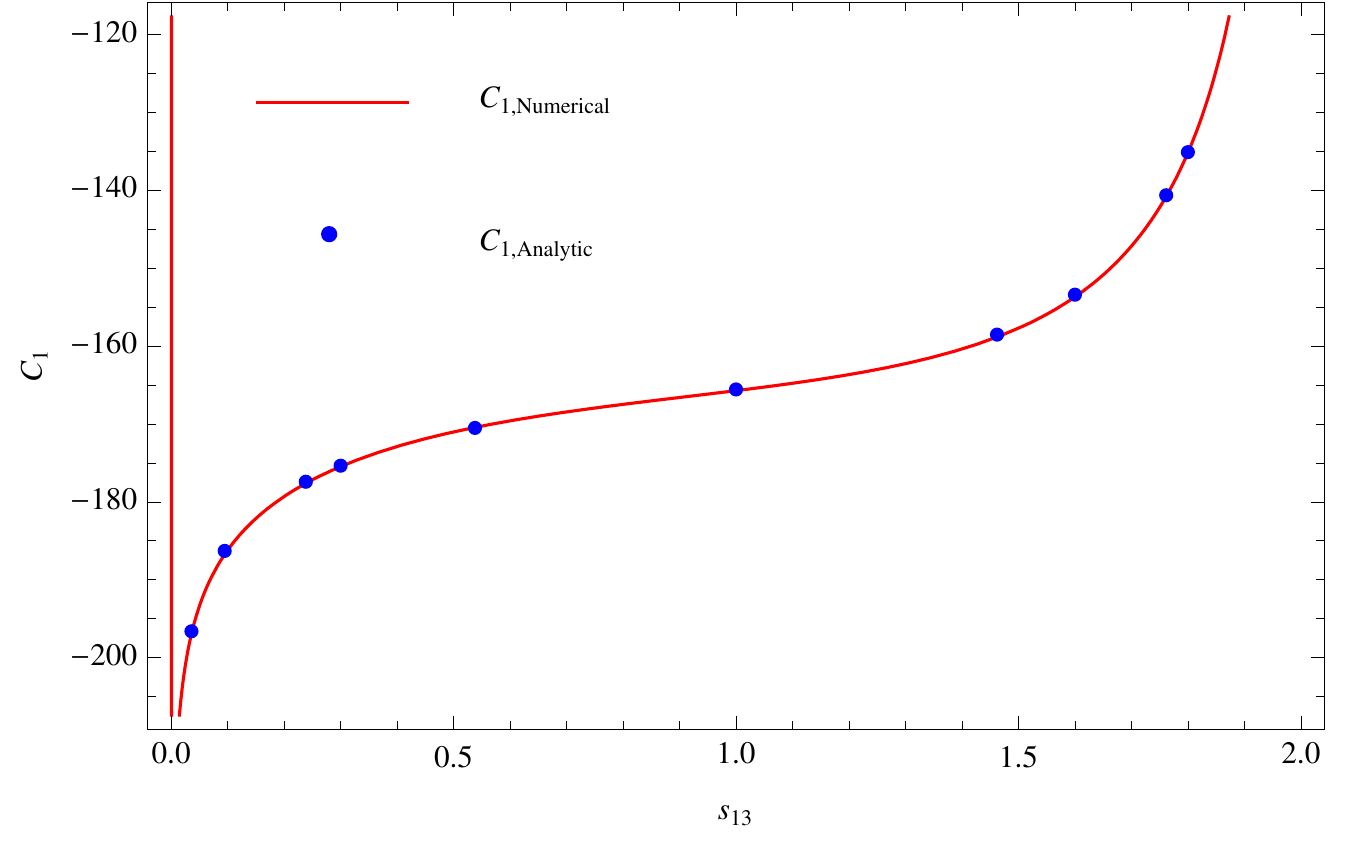}
\end{minipage}
\end{center}
\vspace{-0.5cm}
\caption{Comparison between the analytic and numerical calculation of the coefficients of $L^n$ ($n=4,3,2,1$) for the non-Abelian contribution to the soft function. The red solid lines represent our direct NNLO calculation, and the blue dots are predicted by expanding the resummed results to ${\cal O}(\alpha_s^2) $.} \label{fig:pp-non-abelian}
\end{figure}

\begin{figure}[!h]
\begin{center}
\begin{minipage}[b]{0.46\linewidth}
\includegraphics[width=\textwidth,angle=0]{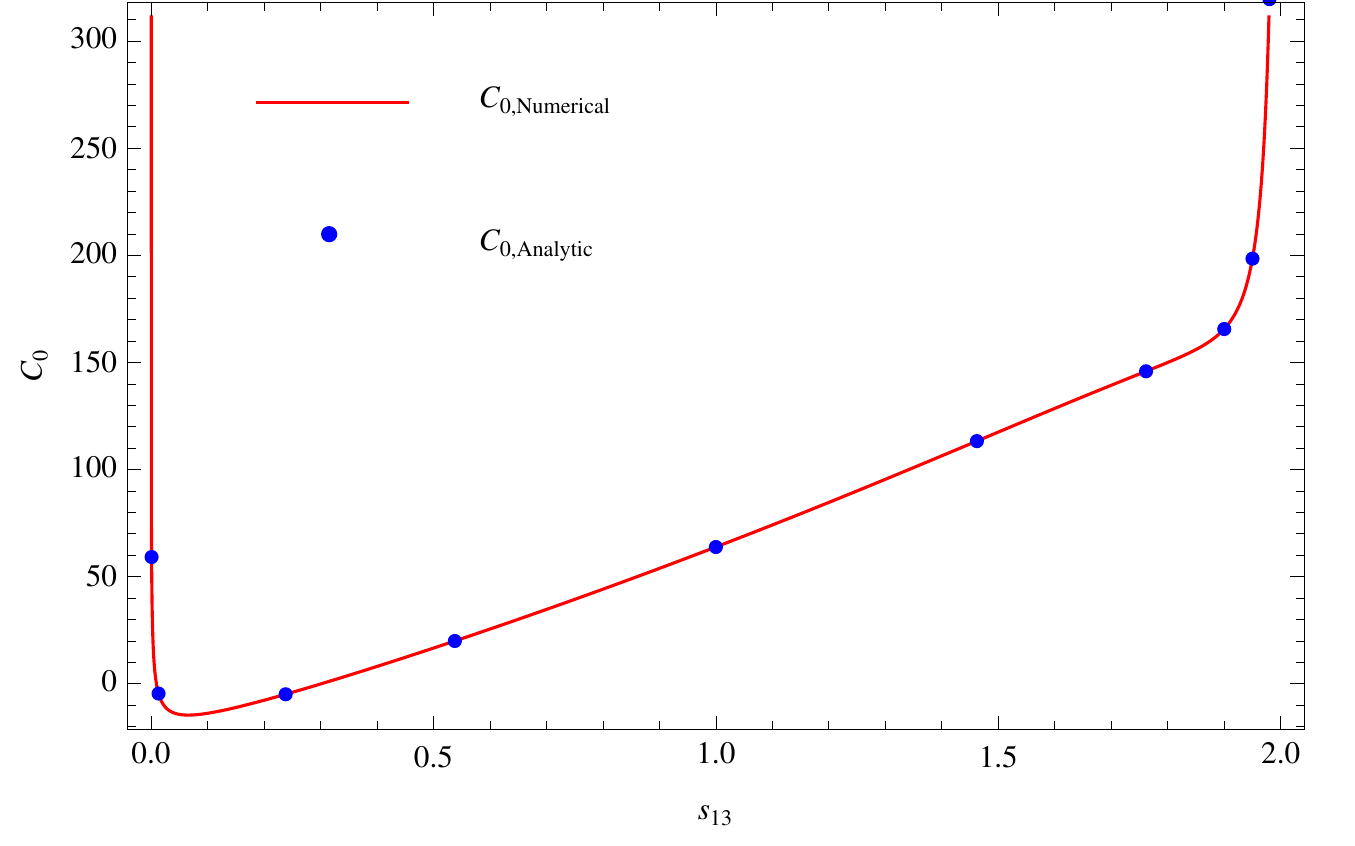}
\end{minipage}
\hspace{0.5cm}
\begin{minipage}[b]{0.46\linewidth}
\includegraphics[width=1.01\textwidth,angle=0]{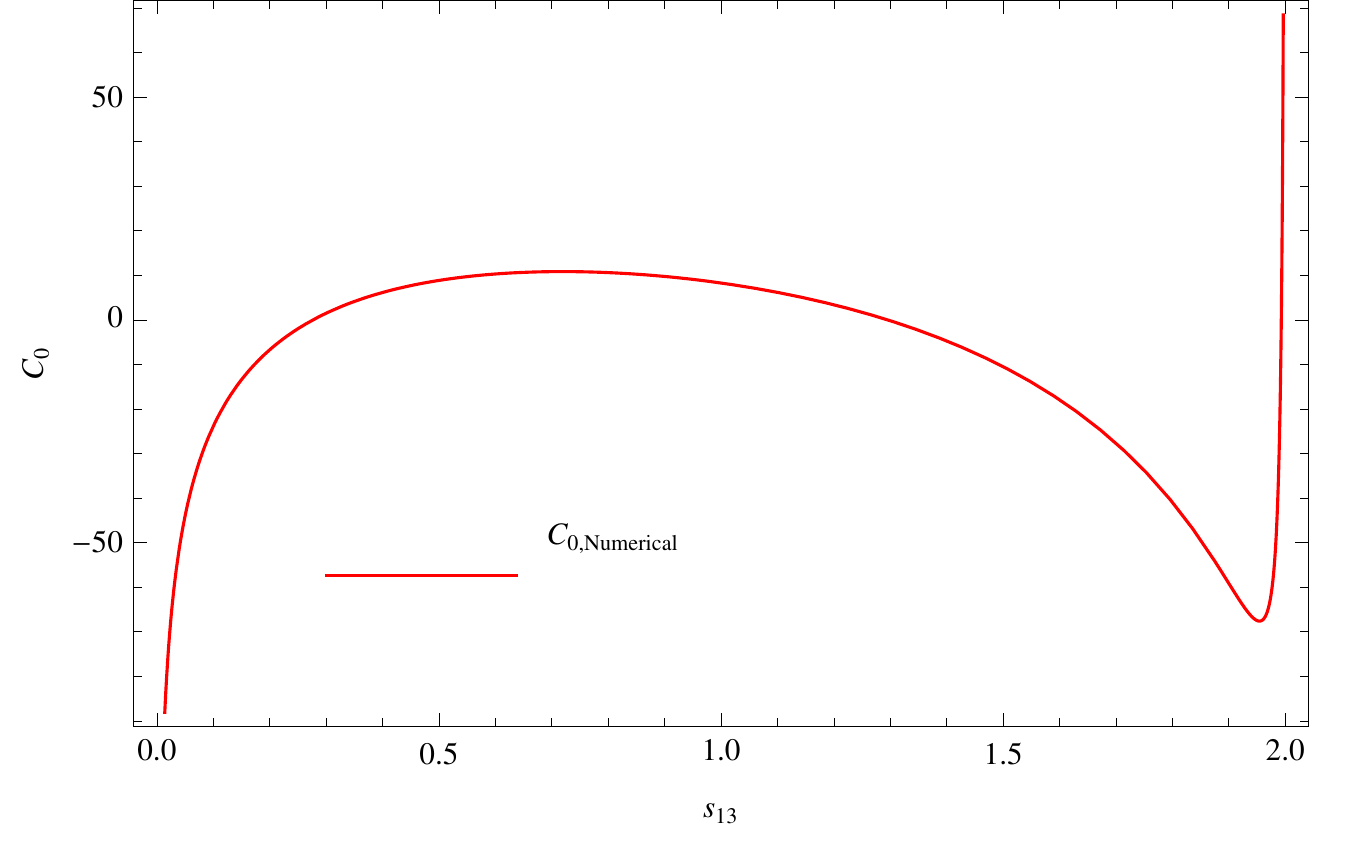}
\end{minipage}
\end{center}
\vspace{-0.5cm}
\caption{$L^0$ contributions to the ${\cal O}(\alpha_s^2)$ soft function, for both 
Abelian (left panel) and non-Abelian (right panel) cases. The red solid lines represent our direct NNLO calculation.  In the Abelian case, the blue dots are obtained by expanding the exponentiated NLO soft function.} \label{fig:pp-L0}
\end{figure}

In Figs.~\ref{fig:pp-abelian}~and~\ref{fig:pp-non-abelian} we compare the coefficients $C_n$ of $L^n$ for $n = 4,3,2,1$ from the direct NNLO computation and the expanded resummed soft function to order $\alpha_s^2$ for the Abelian and non-Abelian contributions.
We find complete agreement between the fixed-order calculations and the resummation predictions.  In fig.~\ref{fig:pp-L0} we show our calculations of $C_0$ for both the Abelian and non-Abelian terms. We note that the Abelian result can be predicted from exponentation of soft emissions.  This prediction is shown in Fig~\ref{fig:pp-L0}, and agrees perfectly with our numerical derivation.  The non-Abelian contribution to the $C_0$ term is new.


\section{Conclusions} \label{sec:conc}

In this manuscript we have presented a calculational framework for obtaining NNLO results for the soft functions which are ubiquitous in effective field theory descriptions of scattering processes.  Using the approach of sector decomposition, which has long been a crucial part of the fixed-order QCD calculational toolbox, we have extracted all singularities from the soft-function integrands and reduced the computation to a set of numerical integrals which are simple to evaluate.  We have illustrated our technique using the $N$-jettiness event shape variable~\cite{Stewart:2010tn}.  Several recent ideas suggested in the literature require the NNLO soft function for $N$-jettiness.  These include the improved theoretical description of electron-nucleus collisions~\cite{Kang:2013wca} and deep inelastic scattering~\cite{Kang:2013nha}, and a recently proposed NNLO subtraction scheme for LHC processes containing final-state jets~\cite{wjet}.  We have shown numerical results for the one-jettiness soft function in both $ep$ and $pp$ collisions that address these needs.  

Although we have focused on $N$-jettiness as an example, the techniques introduced hold more generally for SCET$_{\text{I}}$ observables which feature constraints on the light-cone momenta of the final-state radiation.  In such cases all singularities that appear in the integrand can be regulated with dimensional regularization.  For SCET$_{\text{II}}$ observables which instead have a constraint on the transverse momentum of the measured radiation, additional singularities not controlled by dimensional regularization appear.  We believe that our formalism can be extended to also provide NNLO calculations in such cases when an appropriate regulator is chosen~\cite{Chiu:2011qc,Becher:2011dz,Chiu:2012ir}.  We look forward to pursuing these and other extensions of our work.

\section{Acknowledgments}
The work of R.~B. is supported by the U.S.\ Department of Energy, Division
of High Energy Physics, under contract DE-AC02-06CH11357.  The work of X.~L. is supported by the U.S.\ Department of Energy.  
The work of F.~P. is supported by the U.S.\ Department of Energy, Division of High Energy Physics, under contract DE-AC02-06CH11357 and the grant DE-FG02-91ER40684. 

\section*{Appendix}

We present here the analytic result for the one-jettiness soft function in electron-proton collisions.  We organize this result according to color structure, and to whether it arises from a double-real or a real-virtual contribution:
\begin{equation}
S^{(2)}({\cal T}_1) = C_F \,C_A \, S^{(2)}_{RR,C_FC_A}({\cal T}_1) +C_F \,N_F \,T_R \,S^{(2)}_{RR,N_F}({\cal T}_1) +S^{(2)}_{RV}({\cal T}_1). 
\end{equation}
The individual components are as follows:
\bea
S^{(2)}_{RR,C_FC_A}({\cal T}_1) &=& \left[
\frac{2}{\epsilon^3}
+ \frac{1}{\epsilon^2} \left(
\frac{11 }{3}
+ 4  L_{12}
\right) \,
+ \frac{1}{\epsilon} \left(
\frac{67}{9} - \frac{4 \pi ^2}{3}
+ 4  L_{12}^2 + \frac{22}{3} L_{12}
\right) \right. \nn \\
&& \left.  \hspace{5.ex}
+  
\left(
\frac{404}{27}
-\frac{11 \pi ^2}{6}
-\frac{58 \zeta_3}{3}
+
\frac{8}{3} L_{12}^3
+\frac{22}{3}L_{12}^2
+ \left( \frac{134}{9} -\frac{8}{3} \pi ^2 \right) L_{12}
\right) \right. \nn \\
&& \left.  \hspace{5.ex} + 
\epsilon \left(
\frac{4 L_{12}^4}{3}
+\frac{44 L_{12}^3}{9}
+ \left( \frac{134 }{9}
-\frac{8}{3} \pi ^2 \right)  L_{12}^2
+
\left(
\frac{808 }{27}
-\frac{11 \pi ^2 }{3}
-\frac{116  \zeta_3}{3} \right) L_{12} \right. \right. \nn \\
&& \left. \left.  \hspace{10.ex}
+\frac{2140}{81}
+\frac{335 \pi ^2}{54}
-\frac{682 \zeta_3}{9}
-\frac{17 \pi ^4}{36}
\right)
\right]    {\cal T}_1^{-1-4\epsilon}, 
\eea
\bea
S^{(2)}_{RR,N_F}({\cal T}_1) &=& 
\left[
\frac{-4}{3 \epsilon^2} \, 
- \frac{1}{\epsilon} \left(
\frac{20}{9}
+ \frac{8 }{3 } \, L_{12}
\right) \,
-\frac{112}{27}
+\frac{2 \pi ^2}{3}
-\frac{8}{3}  L_{12}^2-\frac{40 }{9} \, L_{12} \right. \nn \\
&& \left.  + 
\epsilon  \, 
\left( 
-\frac{16}{9}  L_{12}^3
-\frac{40 L_{12}^2}{9}
+ 
\left( \frac{4 \pi ^2 }{3}
-\frac{224}{27} \right) \, L_{12}
+\frac{248 \zeta_3}{9}
-\frac{74 \pi ^2}{27}
-\frac{80}{81}
\right) 
\right] \, {\cal T}_1 ^{-1-4\epsilon}, \nn \\
\eea
\bea
S^{(2)}_{RV}({\cal T}_1)&=&C_A\,C_F \left[
\frac{-2}{\epsilon^3} 
- \frac{4}{\epsilon^2} L_{12}\,
+ \frac{1}{\epsilon} \left( 
\pi ^2-4 L_{12}^2
\right) 
+\frac{16 \zeta_3}{3}
-\frac{8}{3}  L_{12}^3+2 \pi ^2 L_{12} \right.
\nn \\ 
&&\left.
+ \epsilon \left(
-\frac{4}{3}  L_{12}^4+2 \pi ^2 L_{12}^2
+ \frac{32 \zeta_3}{3} L_{12}
-\frac{\pi ^4}{60}
\right)
\right] \,  {\cal T}_1^{-1-4\epsilon} .
\eea
We have set $L_{ij} = \text{log}(n_i \cdot n_j/2)$.

\end{document}